\documentclass[aps,prd,twocolumn,superscriptaddress,nofootinbib,floatfix]{revtex4-2}

\makeatletter

\let\supp@l@section\l@section
\let\supp@l@subsection\l@subsection

\long\def\supp@gobble#1#2{}

\newcommand{\HideEarlierTOCEntries}{\let\l@section\supp@gobble
  \let\l@subsection\supp@gobble
}

\newcommand{\ShowSupplementTOCEntries}{\let\l@section\supp@l@section
  \let\l@subsection\supp@l@subsection
}

\makeatother

\usepackage[
  colorlinks=true,
  linkcolor=blue,
  citecolor=blue,
  urlcolor=blue
]{hyperref}

\usepackage{amsmath,amssymb,bm}
\usepackage{graphicx}
\usepackage{xcolor}
\usepackage[normalem]{ulem}

\usepackage{lineno}
\usepackage{comment} 

\usepackage{appendix}

\newcommand{\di}{\text{d}}
\newcommand{\pare}[1]{\left(#1\right)}
\newcommand{\parea}[1]{\left[#1\right]}
\newcommand{\pareb}[1]{\left\{#1\right\}}
\newcommand{\avg}[1]{\left<#1\right>}
\newcommand{\abs}[1]{\left|#1\right|}

\newcommand{\os}{\omega_0}

\newcommand{\nf}{N_\text{flips}}

\newcommand{\qmin}{\ensuremath{q_0}}

\newif\ifshowrevisions
\showrevisionstrue
\showrevisionsfalse

\ifshowrevisions
 \newcommand{\cut}[1]{\textcolor{gray}{\sout{#1}}}
 
\else
  \newcommand{\cut}[1]{}
  
\fi

\begin{document}

\title{Quantum Observables of Nonperturbative Superradiant Scattering}

\author{A. Arvanitaki}
\email{aarvanitaki@perimeterinstitute.ca}
\affiliation{Perimeter Institute for Theoretical Physics, 31 Caroline St N, N2L 2Y5, Waterloo, ON, Canada}
\author{S. Dimopoulos}
\email{savas@stanford.edu}
\affiliation{Leinweber Institute for Theoretical Physics, Stanford University, 382 Via Pueblo Mall, 94305, CA, USA}
\affiliation{Perimeter Institute for Theoretical Physics, 31 Caroline St N, N2L 2Y5, Waterloo, ON, Canada}
\author{M. Galanis}
\email{mgalanis@stanford.edu}
\affiliation{Leinweber Institute for Theoretical Physics, Stanford University, 382 Via Pueblo Mall, 94305, CA, USA}
\author{D. J. Li}
\email{djpli@stanford.edu}
\affiliation{Leinweber Institute for Theoretical Physics, Stanford University, 382 Via Pueblo Mall, 94305, CA, USA}
\affiliation{SLAC National Accelerator Laboratory, Stanford University, Menlo Park, CA 94025, USA}

\begin{abstract}
Inelastic scattering from $N$ coherent spins, or any two-state system, scales as $N^2$, invalidating the Born approximation at large $N$ even at weak coupling. We develop a nonperturbative theory of this superradiant scattering and its backreaction on the spins. Since one particle can exchange many quanta, the target’s energy-transfer and energy-variance rates keep their favorable $N^2$ scaling even after the cross section saturates, unless a conservation law limits each encounter to exchange of one quantum. We apply the theory to millicharged-particle searches, as well as neutron and charged-particle beams that can test quantum protocols for detecting cosmic relics.
\end{abstract}

\maketitle

\textbf{Introduction---}
Collective coherence enhances photon emission and scattering through Dicke superradiance~\cite{Dicke1954} and superradiant Raman scattering~\cite{Schneble2004}. We recently generalized this principle to the inelastic scattering of arbitrary particles~\cite{Arvanitaki:2024taq,Galanis:2025amc}: $N$ two-level systems in identical coherent superpositions can be excited or de-excited at rates proportional to $N^2$. Coherence constrains the \textit{momentum transfer}, i.e. $qR\lesssim1$, rather than  the \textit{incident momentum} $k$, allowing large targets with $R\gg k^{-1}$ and consequently large interactions rates. This opens prospects for detecting dark matter and the cosmic neutrino background~\cite{Arvanitaki:2024taq}.

The Born approximation can fail at large $N$ even for weak microscopic couplings. At mean-field level the two-level system ensemble behaves as a classical precessing dipole~\cite{Arecchi1972}, producing a potential analogous to that encountered by light in a time-modulated dielectric~\cite{Morgenthaler1958,ZuritaSanchez2009}. We develop a nonperturbative treatment of scattering off this potential using the time-dependent eikonal approximation~\cite{Glauber:eikonal,Liu2024}. This approximation follows nearly straight particle trajectories while retaining the accumulated scattering phase without requiring it to be small. 

We retain the quantum fluctuations of the spin ensemble around the
classical precessing dipole and use them to compute the backreaction of
nonperturbative scattering on the target. This gives access to quantum
observables, in particular the mean energy transfer and the energy
diffusion (growth of the energy variance) of the target, which we
previously identified as coherently enhanced~\cite{Arvanitaki:2024taq,
Galanis:2025amc}. We apply the formalism to \emph{spin-orbit}
interactions, in which the target spins couple to the projectile's motion
and can exchange any number of quanta with it, and to \emph{spin-spin}
interactions, in which conservation of the total spin projection limits a
spin-$1/2$ projectile to a single quantum per encounter. In the former
case, multiquantum exchange, permits the observables to retain their Born scaling
even after the total cross section saturates; in the latter, they saturate
together with the cross section, at the geometric rate for contact
interactions.

Our framework covers both contact and long-range inelastic interactions with the target's internal degrees of freedom (spins). Polarized thermal neutrons and charged-particle beams offer laboratory tests of collective scattering and quantum protocols for relic detection, while terrestrial millicharged particles provide an illustrative cosmic relic search.

\textbf{Coherent inelastic scattering---}
Consider $N$ noninteracting two-level systems, each prepared in the same coherent superposition,
\begin{equation}
|\psi_i(t)\rangle
=c_g|g\rangle+c_e e^{-i\omega_0t}|e\rangle,
\qquad |c_g|^2+|c_e|^2=1,
\label{eq:prod_state}
\end{equation}
where $|g\rangle$ and $|e\rangle$ are split by $\omega_0$. Since two-level systems generally admit a spin-$1/2$ representation, we will refer to such a system simply as spin. The mean spin precesses like a classical dipole, with $\langle s_+^{(i)}(t)\rangle=c_e^*c_g e^{i\omega_0t}$ and constant $\langle s_z^{(i)}\rangle=(|c_e|^2- |c_g|^2)/2$. The collective spin is $\vec J=\sum_i\vec s_i$. 

A scattering particle $P$ couples linearly to the target spins through
\begin{equation}
\mathcal H_i(\mathbf r)=g_i(\mathbf r)
\left[\mathcal O_-s_+^{(i)}+\mathcal O_+s_-^{(i)}
+\mathcal O_zs_z^{(i)}\right],
\label{eq:generalHi}
\end{equation}
where $s_\pm=s_x\pm i s_y$, $\mathcal O_\alpha$ act on $P$, and $g_i(\mathbf r)$ specifies the spatial coupling. In the semiclassical approximation, we replace the spins by their expectation values in Eq.~\eqref{eq:prod_state}, whose regime of validity will be discussed below.

The transverse amplitude is maximal for an equatorial coherent spin state with $c_g=c_e=1/\sqrt{2}$, giving $\langle s_z^{(i)}\rangle=0$ and $\langle s_\pm^{(i)}(t)\rangle=e^{\pm i\omega_0t}/2$. Summing over the target's constituents gives the periodic potential acting on the projectile P
\begin{equation}
V(\mathbf r,t)=V_+(\mathbf r)e^{-i\omega_0t}
+V_-(\mathbf r)e^{i\omega_0t},
\label{eq:potential}
\end{equation}
where $V_\pm$ may retain operators acting on $P$'s internal space, and by energy conservation, they excite or de-excite $P$ by energy $\omega_0$, respectively. At small momentum transfer, $|V_\pm|\propto N$, yielding the $N^2$ enhancement in the Born approximation~\cite{Arvanitaki:2024taq}.

As $N$ increases, the Born approximation can fail, but if $|V|\ll E_{\rm kin}\equiv mv^2/2$, and $k^{-1}$ is short compared to the length scales of the interaction, scattering remains forward-peaked. This permits us to apply the eikonal approximation to $P$'s wavefunction. 

\textbf{Floquet scattering---}For $V_\pm$ independent of the projectile's spin, which we will refer to as \emph{spin-orbit} scattering, we can employ Floquet theory \cite{Floquet1883, Shirley1965, Howland1979} 
    for a time-periodic potential and apply the eikonal approximation \cite{Glauber:eikonal, Liu2024} to compute the cross section of transferring $n\omega_0$ of energy to $P$ to leading-order (LO) in $|V/E_{\rm kin}|$ and $1/(kL)$:
\begin{align}
\sigma_n^{\rm LO}
&=\int d^2\mathbf b\,
\left|\delta_{n0}-J_n[\bar\chi(\mathbf b)]\right|^2,
\label{eq:eikonal:sigma_n}\end{align}
where $\mathbf b$ is the impact parameter, $J_n$ is a Bessel function, $k=mv$ is $P$'s incident momentum. This result holds at leading order (LO) in $|V/E_{\rm kin}|$ and $(kL)^{-1}$, where $L$ is the spatial variation scale of $V$. The total cross section is $\sigma_{\rm tot}^{\rm LO}=2\int d^2\mathbf b\,
\left[1-J_0[\bar\chi(\mathbf b)]\right]$ and $\bar{\chi}$ is the amplitude of the eikonal phase:
\begin{equation}
\bar\chi(\mathbf b)=\frac{2}{v}
\left|\int_{-\infty}^{\infty}dz\,
V_+(\mathbf b,z)e^{-iq_0z}\right|,
\qquad q_0\equiv \frac{\omega_0}{v}.
\label{eq:chiLO_asympt}
\end{equation}
Here $q_0$ is the minimum single-quantum momentum transfer. These expressions also assume $|n|\omega_0\ll E_{\rm kin}$ in the populated channels so that deflection is small. The Born series follows by expanding under $\bar\chi\ll 1$, while $\bar\chi\gtrsim1$ permits multi-quanta exchange with characteristic $|n|\sim\bar\chi$, due to the appreciable weight of $J_n(\bar\chi)$ up to that order.

\begin{figure}[!t]
\centering
\includegraphics[width=\columnwidth]{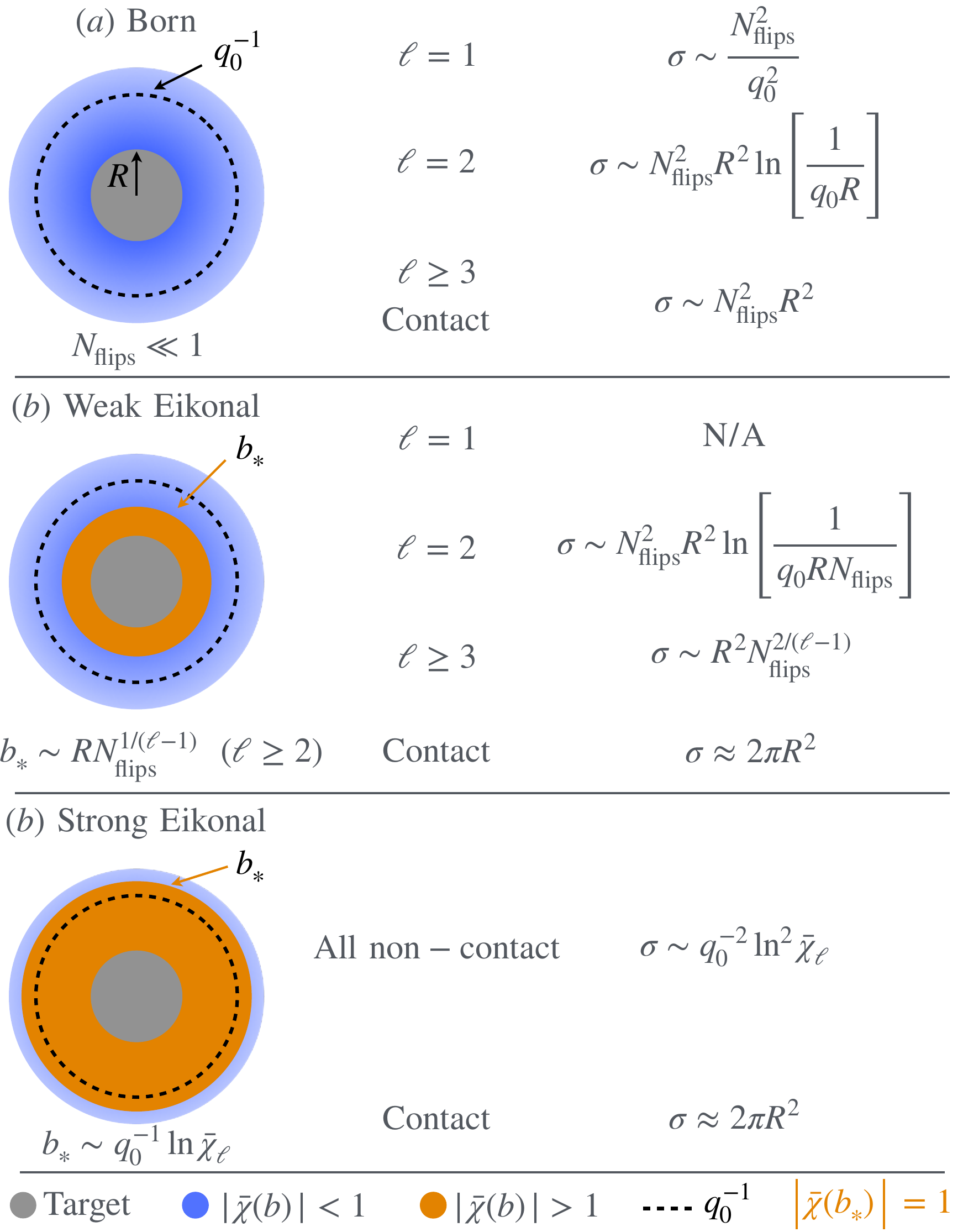}
\caption{Born, weak-eikonal, and strong-eikonal regimes at $q_0R\ll1$, with coupling increasing from top to bottom. Gray denotes the target; orange and blue indicate strong and weak scattering, separated by $|\bar\chi(b_*)|=1$. Scalings apply to $V_\ell(r)=V_0(R/r)^\ell$ outside the target, while contact denotes $V_0\Theta(R-r)$. Born scattering holds at every impact parameter only if $\nf\ln[1/(q_0R)]\ll1$. Only the contact saturation prefactor is retained.}
\label{fig:scattering_regimes}
\end{figure}

Contact scattering saturates at $2\pi R^2$, but long-range interactions also contributes at $b\gg R$. For $V_\ell(r)=V_0(R/r)^\ell$ at $r\geq R$ and $q_0R\ll1$, the phase decays exponentially beyond $b\sim q_0^{-1}$, acting as an effective mediator mass. Define $b_*$ by $|\bar\chi(b_*)|=1$. The integrand $1-J_0(\bar\chi)$ is of order unity at large phase and $\bar\chi^2/4$ at small phase, giving a strong core and weak tail:
\begin{equation}
\sigma_{\rm tot}^\text{LO}\sim b_*^2+
\int_{b_*}^{\infty}db\,b\,|\bar\chi(b)|^2.
\label{eq:core_tail_scaling}
\end{equation}
The core contribution scales with its area, while the tail is the Born integral with lower cutoff $b_*$. Consider first the \emph{weak-eikonal} regime, defined when $q_0 b_*\ll 1$. For $\ell>1$ and $b\geq R$, $|\bar\chi(b)|\sim\nf(R/b)^{\ell-1}$ and  $b_*\sim R\nf^{1/(\ell-1)}$, where $\nf=2\pi|V_0|R/v$ sets the phase strength. For $\ell=2$, the perturbative tail dominates, yielding $\sigma_{\rm tot}\sim b_*^2\ln[1/(q_0b_*)]$ and the total cross-section remains near the Born approximation value despite non-perturbative scattering near the target. For $\ell\geq3$, both terms in Eq.~\eqref{eq:core_tail_scaling} scale as $b_*^2$. 

In the \emph{strong eikonal} regime, $b_*\gg q_0^{-1}$ enters the Bessel function's tail. Define $\bar\chi_\ell=c_\ell\nf(q_0R)^{\ell-1}$ with $c_\ell=2^{(3-\ell)/2}/[\sqrt\pi\,\Gamma(\ell/2)]$. For $\bar\chi_\ell\gg1$, all power laws, including $\ell=1$, give a universal cross-section, up to logarithms
\begin{equation}
b_*\sim q_0^{-1}\ln\bar\chi_\ell,
\qquad\sigma_{\rm tot}\sim q_0^{-2}\ln^2\bar\chi_\ell.
\label{eq:strong_tail_scaling}
\end{equation}
Figure~\ref{fig:scattering_regimes} summarizes these regimes.

\textbf{Internal degrees of freedom---}
For operator-valued $V_\pm$, the scalar eikonal factor becomes a path-ordered matrix exponential. Consider a $P$ with spin-$1/2$ and a contact interaction at energy scale $\Lambda$

\begin{equation}
H_{\rm int}=\frac{1}{\Lambda^2}\sum_i
\vec s_P\cdot\vec s_i(t)\,\delta^{(3)}(\mathbf r-\mathbf r_i).
\label{eq:spinspin_internal}
\end{equation}
Conservation of $J_z+s_{P,z}$ allows only one inelastic channel per incident polarization: a spin-up (spin-down) probe loses (gains) one quantum. For a uniform equatorial target of density $n_s$, define $V_0=n_s/(2\Lambda^2)$ and $\Omega^2=V_0^2+\omega_0^2$. Both polarizations give
\begin{equation}
\sigma_{\rm flip}\simeq\frac{2\pi V_0^2}{\Omega^2}
\int_0^R b\,db\,\sin^2\!\left(\frac{\Omega\sqrt{R^2-b^2}}{v}\right).
\label{eq:sigma_flip}
\end{equation}
This is equivalent to a Rabi pulse of strength $V_0$, detuning $\omega_0$, applied for the transit time $2\sqrt{R^2-b^2}/v$. For $|V_0|R/v\gg1$ and $\omega_0R/v\ll1$, $\sigma_{\rm flip}\simeq\pi R^2/2$, while $\sigma_{\rm tot}\simeq2\pi R^2$ includes elastic scattering. A calculation is provided in the \emph{Supplemental Material}. As we will see, conservation of total spins has important consequences for the detector backreaction and observables.

\textbf{Detector backreaction dynamics---}The preceding semi-classical approximation holds in the limit of weak backreaction to the target. To quantify this, we retain the dynamical fluctuations of the spins, by introducing Holstein--Primakoff bosons~\cite{Holstein1940}
\begin{equation}
s_x^{(i)}=\frac12-n_i,\qquad s_y^{(i)}\simeq\frac{i}{2}(a_i-a_i^\dagger),\qquad
s_z^{(i)}\simeq\frac12(a_i+a_i^\dagger),
\label{eq:HP}
\end{equation}
with $[a_i,a_j^\dagger]=\delta_{ij}$ and $n_i=a_i^\dagger a_i$. The initial state is the bosonic vacuum, which retains the transverse projection noise $\operatorname{var}[J_z]=N/4$, and the boson occupation $n_i$ measures loss of $x$-polarization. The linearity of Eq.~\eqref{eq:HP} assumes $\avg{n_i}\ll 1$, or $\avg{\delta J_i}\ll N$ for the collective spins.

\begin{figure}[!t]
    \centering
    \includegraphics[width=\linewidth]{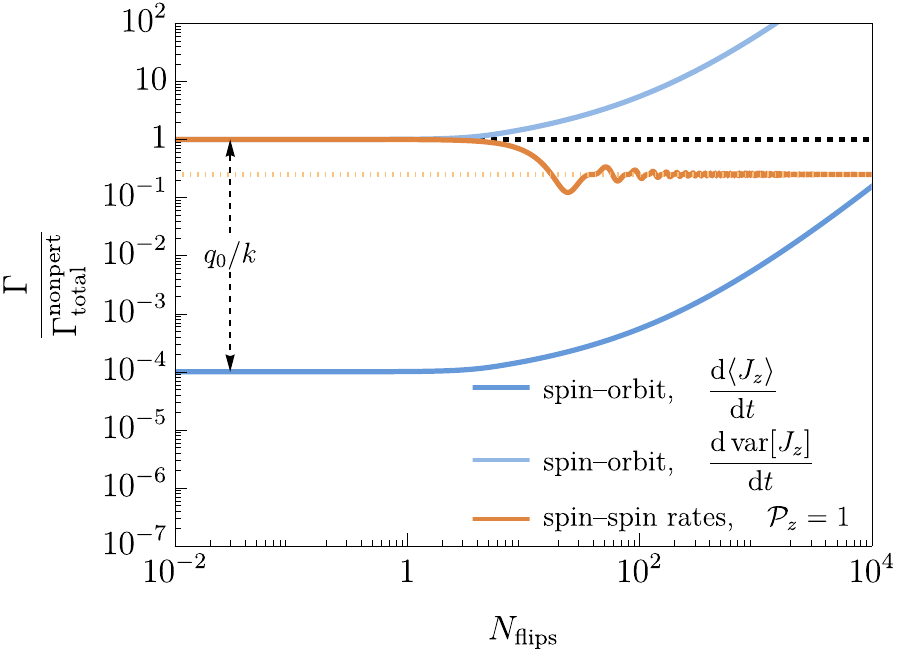}
    \caption{Mean energy-transfer and decoherence rates divided by the total interaction rate, for a monoenergetic bath. The total rate is given by $\Phi\sigma_\text{tot}$, where $\Phi$ is the flux and the total cross-section is given by Eq.~\eqref{eq:core_tail_scaling}. Spin-orbit scattering (blue) permits multiquantum transfer and so Born-like growth with $N_\text{flips}$, while polarized spin-$1/2$ exchange (orange, $\mathcal P_z=1$) approaches the geometric ratio $1/4$. Here $q_0/k=10^{-4}$ and we terminate the plot at $N_\text{flips}\sim k/q_0$, where eikonal approximation breaks down.}
    \label{fig:sigma_born vs eikonal}
\end{figure}

Retaining only linear terms, Eq.~\eqref{eq:generalHi} becomes
\begin{equation}
\begin{split}
\mathcal H_{I,i}\simeq \frac{1}{2}g_i(\mathbf r)\Bigl\{
&\left[\mathcal O_{I,-}e^{i\os t}
+\mathcal O_{I,+}e^{-i\os t}\right]
\\
&+\left[\mathcal O_{I,+}e^{-i\os t}
-\mathcal O_{I,-}e^{i\os t}\right](a_i-a_i^\dagger)
\\
&+\mathcal O_{I,z}(a_i+a_i^\dagger)
\Bigr\},
\label{eq:HP_hamiltonian}
\end{split}
\end{equation}
where $\mathcal O_{I,\alpha}$ are particle operators in the interaction picture. The first line reproduces the classical potential and dresses $P$'s wavefunctions, which, in turn, drive the target's quantum response in the second and third lines.

\textbf{Observables---}The coherently enhanced observables are the energy transfer and variance~\cite{Arvanitaki:2024taq,Galanis:2025amc}. We assume independent incident particles and average over the target's fast precession. LO and NLO refer to the order retained in the eikonal expansion.

For spin-orbit scattering, the mean response before precession averaging is
\begin{equation}
\frac{\di\langle J_z\rangle}{\di t}
=-2n_p\left\langle\int\di^3r\,
e^{-2\operatorname{Im}\chi}
\operatorname{Im}\!\left[V_+(\mathbf r)e^{-i\os t}\right]
\right\rangle_f,
\label{eq:jz_gen}
\end{equation}
where $\langle u\rangle_f=n_p^{-1}\int\di^3k\,f(\mathbf k)u/(2\pi)^3$ averages over an incident distribution of number density $n_p$. At LO, $\chi$ is real: excitation and de-excitation are equally probable, so individual particles exchange energy while the mean transfer vanishes on average.

The first nonzero mean comes from the imaginary NLO correction, which is small, $|\text{Im}\chi^{\rm NLO}|\sim V/E_\text{kin}\ll 1$. We find (see \emph{Supplemental Material}, Secs.~\ref{app:observables} and~\ref{app:NLO_eikonal})

\begin{equation}
\left.\frac{\di\langle J_z\rangle}{\di t}\right|_{\rm NLO}
\simeq\frac{n_p}{m}\left\langle
\int\frac{\di^2q_\perp}{(2\pi)^2v^2}
\left.\partial_q
\bigl[Q^2|\widetilde V_+(\mathbf Q)|^2\bigr]
\right|_{q=q_0}
\right\rangle_f,
\label{eq:jz}
\end{equation}
where $\mathbf Q=(\mathbf q_\perp,q)$. This is equivalent to the first channel-moment $n_p\sum_n n\langle\sigma_n^{\rm NLO}v\rangle_f$, as expected from energy conservation.

In contrast, energy fluctuations are driven by either direction of energy transfer. We find 

\begin{equation}
\left.\frac{\di\operatorname{var}[J_z]}{\di t}\right|_{\rm LO}
\simeq n_p\left\langle km
\int\frac{\di\Omega_{k'}}{4\pi^2}
\sum_{\eta=\pm}|\widetilde V_\eta(\mathbf q_\eta)|^2
\right\rangle_f,
\label{eq:jz2}
\end{equation}
where $\mathbf q_\pm=\mathbf k-\mathbf k'_\pm$ and $k'_\pm=\sqrt{k^2\pm2m\os}$. We approximate $k'_\pm\simeq k$ in the prefactor while retaining energy transfer in the potential.  This is equivalent to the second channel moment $n_p\sum_n n^2\langle\sigma_n^{\rm LO}v\rangle_f$. Proofs of the above equations can be found in the \emph{Supplemental Material}, Secs.~\ref{app:observables} and~\ref{app:decoherence_S}. The same $n^2$ weighting also broadens the final energy spread of $P$.

Remarkably, both rates retain the quadratic Born form. For the mean, the imaginary NLO correction is linear in the potential, so its product with $V_+$ gives a quadratic response. For variance, the corresponding LO channel moment obeys the exact identity

\begin{equation}
\sum_n n^2\sigma_n^{\rm LO}
=\frac12\int\di^2b\,\bar\chi^2
=2\sigma_\pm^{\rm Born},
\label{eq:jz2_born_identity}
\end{equation}
where $\sigma_\pm^{\rm Born}$ is either of the one-quantum Born cross sections. Equation~\eqref{eq:jz2_born_identity} can also be understood as a typical flip of $N_\text{flip}\sim |\bar{\chi}|$ spins per passage. In other words, the eikonal phase sets the characteristic spin transfer, as long as $|n|\os\sim N_\text{flip}\os\ll E_\text{kin}$.

For spin-$1/2$ exchange, conservation of total spin projection instead restricts $n$ to $0,\pm1$. Incident polarization $|\mathcal P_z|\leq 1$ biases the direction of exchange, giving the LO rates
\begin{equation}
\left.\left(\frac{\di\langle J_z\rangle}{\di t},
\frac{\di\operatorname{var}[J_z]}{\di t}\right)\right|_{\rm LO}
\simeq n_p\left\langle(\mathcal P_z,1)\sigma_{\rm flip}^{\rm LO}v\right\rangle_f.
\label{eq:rates_spin}
\end{equation}
For $\mathcal P_z=1$, the rates coincide because every spin-flip event excites the target by one quantum.

Saturation of the scattering rate (up to logs) therefore need not imply saturation of the detector response. Multiquantum exchange can preserve the Born scaling of spin fluctuations and mean energy transfer. On the other hand, conservation laws restricting each encounter to one quantum make both observables inherit the saturation of the total rate. Figure~\ref{fig:sigma_born vs eikonal} compares these behaviors, which motivate the neutron and charged-particle tests below.

\textbf{Polarized thermal-neutron and charged-particle beams---}
Thermal-neutron and charged-particle beams offer laboratory tests of the scattering response and quantum measurement protocols for relic detection. They probe complementary limits. Spin conservation bounds neutron transfer to one quantum per encounter, whereas a charged beam can exchange many quanta through the target's oscillating magnetization. Comparing their mean and variance responses would test the distinction between saturation of the event rate and continued growth of the spin-transfer moments.

Low-energy neutrons interact with polarized $^3$He through
the spin-dependent contact Hamiltonian of
Eq.~\eqref{eq:spinspin_internal}, producing the effective
nuclear field known as pseudomagnetism~\cite{russianpaper}.
Here $\Lambda^{-2}=4\pi a_s/m_r$, where $m_r$ is the reduced
mass and $a_s=(a_t-a_0)/2$ is half the difference between
the triplet and singlet scattering lengths. 
Defining $V_0=2\pi n_s\operatorname{Re}a_s/m_r$, the
rates for independent thermal neutrons with
$\mathcal P_z=1$ and total flux $\Phi$ are
\begin{equation}
\label{eq:neutron:rates}
\begin{aligned}
\frac{\di\langle J_z\rangle}{\di t}
&=\frac{\di\,\operatorname{var}(J_z)}{\di t}\\
&\simeq\frac{\pi\Phi}{2}
\begin{cases}
V_0^2R^4/v_T^2, & |V_0|R/v_T\ll1,\\[2pt]
R^2, & |V_0|R/v_T\gg1.
\end{cases}
\end{aligned}
\end{equation}
Here $v_T=\sqrt{2k_BT_n/m_n}$, with $\langle v^{-2}\rangle_{\rm flux}=v_T^{-2}$ for a Maxwellian incident flux. The two limits describe perturbative growth and geometric saturation, while neglecting absorption.

\begin{figure}[!t]
    \centering
    \includegraphics[width=\columnwidth]{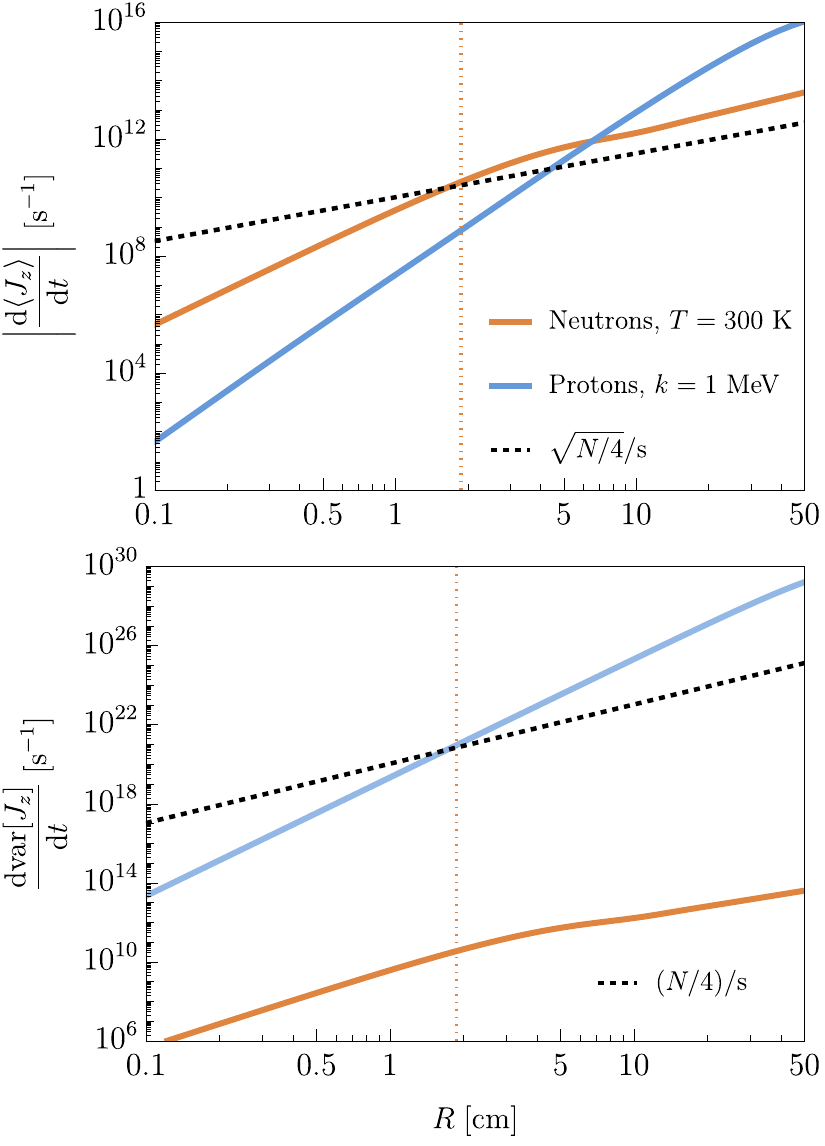}
    \caption{Quantum observables for superradiant scattering of thermal neutrons at 300~K (orange) and an axial proton beam at 1~MeV (blue) with a spin-polarized $^3$He target of radius $R$ in the equatorial product state. The target density is $n_s=10^{20}\,\mathrm{cm}^{-3}$, the neutron flux $\Phi=10^{10}\,\mathrm{cm}^{-2}\mathrm{s}^{-1}$, and the proton flux $\Phi=10^{9}\,\mathrm{cm}^{-2}\mathrm{s}^{-1}$. We assume splittings $\os=10^{-12}$~eV and $10^{-10}~$eV for the neutron and proton plots, respectively, to ensure coherence over all plotted radii $R$. \emph{Top panel:} Mean energy transfer. Neutrons deposit energy, while protons absorb. \emph{Bottom panel:} Variance growth. The black dashed lines correspond to SQL/s, namely $\sqrt{N/4}$/s and $(N/4)$/s, with $N=4\pi n_sR^3/3$, respectively. The transverse extent of the proton beam is matched to the size of the target at every $R$. The vertical dotted orange line separated the Rayleigh-Gans regime from the non-perturbative regime for neutron scattering. Proton scattering is nonperturbative everywhere.}
    \label{fig:beam_rates}
\end{figure}

Abragam and collaborators demonstrated this spin-exchange mechanism by tuning the neutron precession into resonance with the rotating nuclear pseudomagnetic field~\cite{Abragam}. Assuming a $^3$He sample and denoting gyromagnetic ratios as $\gamma$, this detuning is $\Delta_n=(\gamma_{^3\mathrm{He}}-\gamma_{\mathrm{n}})B_{\rm ext} + V_0\avg{s_z}=0$, with $\gamma_{^3\mathrm{He}}=-4.254\,\mu_N$ and $\gamma_{\mathrm{n}}=-3.83\,\mu_N$, where $\mu_N$ is the nuclear magneton, and so resonance is accomplished by a residual axial magnetization $\avg{s_z}$. We note that exact resonance is unnecessary for the rates above, as long as the transit time is short compared to the neutron-target frequency mismatch, $|\Delta_n|R/v_T\ll1$, detuning does not suppress spin exchange.

Charged beams probe the target magnetization through a quasistatic current-dipole potential:
\begin{equation}
\label{eq:beam:potential}
\widetilde V_+(\mathbf Q)
=ieq_bV_s
\frac{\mathbf Q\cdot(\mathbf v\times\mathbf M_+)}{Q^2}
\bigl[F(QR)-F(QR_c)\bigr],
\end{equation}
where $F(u)=3j_1(u)/u$, $V_s=4\pi R^3/3$, $eq_b$ the projectile charge, and $|\mathbf M_+|=|\gamma|n_s/(2\sqrt2)$, with target gyromagnetic ratio $\gamma$. The above potential assumes a uniformly magnetized sphere of radius $R$ enclosed by a concentric superconducting screen of radius $R_c$. 

The initial mean spin-transfer and variance-growth rates follow from Eqs.~\eqref{eq:jz} and~\eqref{eq:jz2}. For nonrelativistic axial beams in the weak-deflection regime, the mean is kinematically suppressed, whereas the variance probes multiquantum exchange already at leading eikonal order.

Figure~\ref{fig:beam_rates} shows the initial mean spin-transfer and variance-growth rates, together with the effective SQL rates corresponding to a variance $(N/4)$/s. We neglect beam attenuation, neutron capture, charged-particle stopping and ionization, and the resulting heating and spin relaxation. Although experimental sensitivity will depend on these effects, readout noise, and measurement statistics, the predicted signals motivate laboratory tests of superradiant scattering and quantum protocols for relic detection.

\begin{figure}[!t]
    \centering
    \includegraphics[width=\columnwidth]{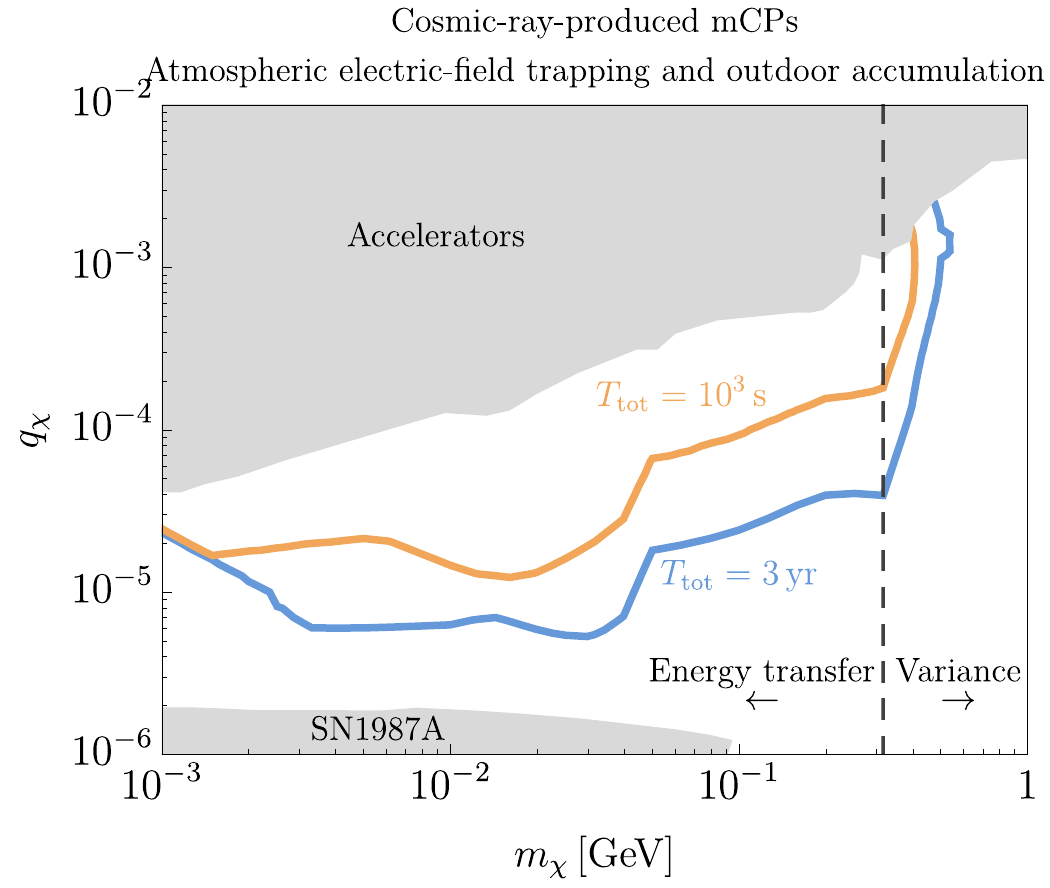}
    \caption{Projected mCP reach for the accumulation scenario of Ref.~\cite{Berlin:2025btf}, with $R=10$~cm, $n_s=3\times10^{21}\,\mathrm{cm}^{-3}$ of $^3$He, and $t_{\rm coh}=10^3$~s. Integration times are $10^3$~s (orange) and $3$~yr (blue). The variance projection further assumes squeezing $10\log_{10}(\xi^{-2})\simeq48$~dB~\cite{Galanis:2025amc}, becoming a more sensitive observable to the right of the dashed line. Gray regions denote accelerator~\cite{Davidson:2000hf,Haas:2014dda,Prinz:1998ua,ArgoNeuT:2019ckq,milliQan:2021lne,ArguellesDelgado:2021lek,PBC:2025sny,Alcott:2025rxn,CMS:2024eyx}
    and SN1987A~\cite{Chang:2018rso} constraints.}
    \label{fig:mCP_reach}
\end{figure}

\textbf{Terrestrial millicharged particles.---} We further study long-range scattering in the eikonal regime with terrestrial millicharged particles (mCPs), which interact with the vector potential of the precessing nuclear-spin target. Cosmic rays generate an irreducible mCP
abundance~\cite{Harnik:2020ugb}, while environmental effects can enhance their local density~\cite{Berlin:2023zpn}. For our projected reach, we adopt the outdoor-accumulator under atmospheric electric-field trapping scenario of Ref.~\cite{Berlin:2025btf}, subject to density-dependent constraints~\cite{Berlin:2025hjs,Budker:2021quh}.

We use the screened potential of Eq.~\eqref{eq:beam:potential} with $q_b=q_\chi$. At fixed $R_c/R$ and for $\omega_0R_c/v\ll1$, the eikonal phase scales as $e|q_\chi\gamma|n_sR^2$, giving the nonperturbative threshold
$|q_\chi|_{\mathrm{crit}}
\sim(e|\gamma|n_sR^2)^{-1}$
up to geometric factors. Our reach is deep within the non-perturbative regime.

For nonrelativistic mCPs thermalized at room temperature $T$, Eqs.~\eqref{eq:jz} and~\eqref{eq:jz2} give the energy-transfer and spin-variance growth rates. Figure~\ref{fig:mCP_reach} shows the potential reach of a $^3$He target in an equatorial coherent spin state, including squeezing and magnification~\cite{Arvanitaki:2024taq,Galanis:2025amc}. Demanding that the internal spin-induced magnetic field not overcome the externally applied one precludes from tuning $\os$ to arbitrarily low values. Taking $|B_{\mathrm{ext}}|=|B_{\mathrm{int}}|=|\gamma|n_s/3$, with $\omega_0=|\gamma B_{\mathrm{ext}}|$ and $v_{\mathrm{th}}=\sqrt{2T/m_\chi}$, coherence requires $\omega_0R/v_{\mathrm{th}}\lesssim1$, or $m_\chi\lesssim\frac{2T}{\gamma^2B_{\mathrm{int}}^2R^2}$.
For the parameters of Fig.~\ref{fig:mCP_reach} and $\gamma=\gamma_{^3\text{He}}$, this allows coherence up to $m_\chi\simeq15$~GeV. This discussion shows how our formalism can be applied to cosmic relic searches. 

\textbf{Prospects---}
The very feature that makes superradiance compelling--interaction rates scaling as $N^2$--also makes a nonperturbative treatment essential at sufficiently large $N$. Here we developped such a framework and applied it to concrete physical examples.

We motivate experimental studies of superradiant scattering using polarized thermal neutrons or proton and electron beams. Such experiments would provide a demonstration of superradiance for massive Standard Model particles while offering a controlled testbed for quantum detection protocols. Using well-understood Standard Model physics, these setups could enable the development and validation of methods for future searches for axions, other dark matter candidates, and ultimately the cosmic neutrino background. 

\textbf{Acknowledgements---} We thank David Cory, Andy Geraci, Kent Irwin,  Mark Kasevich, Maxim Pospelov, Dmitry Pushin, and Nicholas Rapidis for valuable conversations. We also thank Asher Berlin, Zachary Bogorad, Peter W. Graham, and Harikrishnan Ramani for sharing their numerical data from \cite{Berlin:2025btf} with us. This work was supported in part by
NSF Grant PHY-2310429 and the Leinweber Institute for Theoretical Physics at Stanford. MG also acknowledges support by the
Simons Foundation Grant (824870, PG). Research at Perimeter Institute is supported in part by the Government of Canada through the Department of Innovation, Science and Economic Development and by the Province of Ontario through the Ministry of Colleges and Universities. AA is grateful for the support of the Stavros Niarchos Foundation.

\bibliographystyle{apsrev4-2}
\bibliography{biblio}
\clearpage
\onecolumngrid

\setcounter{section}{0}
\setcounter{equation}{0}
\setcounter{figure}{0}

\renewcommand{\thesection}{S\arabic{section}}
\renewcommand{\theequation}{S\arabic{equation}}
\renewcommand{\thefigure}{S\arabic{figure}}

\begin{center}
{\large\bfseries Supplemental Material}\\[4pt]
{\itshape Quantum Observables of Nonperturbative Superradiant Scattering}
\end{center}

\addtocontents{toc}{\protect\ShowSupplementTOCEntries}

\begingroup
\HideEarlierTOCEntries
\def\contentsname{Supplemental Contents}
\tableofcontents
\endgroup

\vspace{1em}
\begin{center}
\rule{0.65\textwidth}{0.5pt}
\end{center}
\vspace{1em}

The sections that follow outline the computations that lead to the results in the main text. The outline is as follows: in Sec.~\ref{app:spin-orbit} we develop the theory of multiquantum superradiant scattering, in Sec.~\ref{app:spin_dep} we study spin-spin scattering, and in Sec.~\ref{app:rates} we compute rates for various applications of our formalism.

\section{Multiquantum scattering}
\label{app:spin-orbit}

The spin-orbit potentials allows for the exchange of, in principle, an arbitrary number of quanta between the scattering particle and the target. In this appendix we apply the eikonal approximation in Sec.~\ref{app:eikonal} to compute $P$'s wavefunctions, and then derive the response of the spin-system to the dressed bath, Sec.~\ref{app:spin_response}. This allows us to compute quantum observables in Sec.~\ref{app:observables}. 

Next, we provide a second independent derivation of our main results using the Floquet-eikonal formalism for scattering off time-periodic potentials. We review Floquet-eikonal scattering in Sec.~\ref{app:Floquet}, provide a general derivation of decoherence using the $S$-matrix in Sec.~\ref{app:decoherence_S}, and compute the NLO corrections to the channel cross-sections in \ref{app:NLO_eikonal}.

\subsection{Eikonal approximation}
\label{app:eikonal}

In first quantization, $P$'s wavefunctions satisfy Schr\"{o}dinger's equation

\begin{equation}
    i\frac{\partial \psi}{\partial t}=-\frac{1}{2m}\nabla^2\psi + V(\mathbf{r},t)\psi,
    \label{eq:schr}
\end{equation}
with $V(\mathbf r,t)=V_+(\mathbf r)e^{-i\omega_0t}
+V_-(\mathbf r)e^{i\omega_0t}.$ Under the conditions outlined in the main text, we make the eikonal ansatz

\begin{eqnarray}
    \psi_E =\frac{1}{\sqrt{V}}e^{-iEt}e^{ikz}e^{i\chi(\mathbf{r},t)},
\end{eqnarray}
with $E=\frac{k^2}{2m}$. Here we chose the incoming momentum axis to be $z$, and so $\mathbf{r}=(\mathbf{b},z)$, where $\mathbf{b}$ is the impact parameter. This $z$-axis is in general misaligned compared to the quantization axis of the spin system.

Equation~\ref{eq:schr} becomes

\begin{equation}
    \pare{\frac{\partial}{\partial t}+v\frac{\partial}{\partial z}}\chi = - V(\mathbf{r},t)  -\frac{1}{2m}\pare{|\nabla\chi|^2-i\nabla^2\chi},
\end{equation}

In the paraxial approximation we may write $\chi \approx \chi_0+\chi_1$, with $\chi_1\ll\chi_0$, which gives

\begin{align}
    \pare{\frac{\partial}{\partial t}+v\frac{\partial}{\partial z}}\chi_0 &\approx -V(\mathbf{x},t)\label{eq:eik_chi0}\\
        \pare{\frac{\partial}{\partial t}+v\frac{\partial}{\partial z}}\chi_1 &\approx -\delta V(\mathbf{b},z,t)\label{eq:eik_chi1},
\end{align}
where we defined

\begin{equation}
    \delta V(t,\mathbf{b},z) = \frac{1}{2m}\parea{\abs{\nabla_{(z,\mathbf{b})} \chi_0(t,z,\mathbf{b})}^2 - i\nabla_{(z,\mathbf{b})}^2\chi_0(t,z,\mathbf{b})}.
\end{equation}

Equations~\eqref{eq:eik_chi0} and~\eqref{eq:eik_chi1} can be solved with the method of characteristics. At LO the eikonal phase is real

\begin{equation}
    \chi_0(t,z,\mathbf{b}) = -\frac{1}{v}\int_{-\infty}^z\di z' \,V\pare{t+\frac{z'-z}{v},z',\mathbf{b}},
    \label{eq:eik_LO}
\end{equation}
while at NLO it receives an imaginary correction

\begin{equation}
\begin{split}
    \text{Im}(\chi_1) = \frac{1}{2mv^2}\int_{-\infty}^z \di z'&\left\{(z-z')\pare{\frac{\os^2}{v^2} - \nabla_\perp^2}\parea{V_+(\mathbf{b},z')e^{-i\os \tau}+V_-(\mathbf{b},z')e^{i\os \tau}}\right.\\
    &\quad-\parea{\frac{\partial V_+(\mathbf{b},z')}{\partial z'}e^{-i\os \tau} + \frac{\partial V_-(\mathbf{b},z')}{\partial z'}e^{i\os \tau}}\\
    &\quad\left.
    -\frac{i\os}{v}
    \parea{V_+(\mathbf{b},z')e^{-i\os \tau}-V_-(\mathbf{b},z')e^{i\os \tau}}\right\},\label{eq:imag_eik}
\end{split}
\end{equation}
where $\tau\equiv t + \frac{z'-z}{v}$ and parametrically $\text{Im}(\chi_1)\sim \frac{V}{E}\ll1$ within the eikonal approximation.

The eikonal solution $\psi_E\approx V^{-1/2} e^{-i(Et-kz)}e^{i[\chi_0+\text{Re}(\chi_1)]}e^{-\text{Im}(\chi_1)}$ corresponds to a free particle at asymptotic past, $z\to -\infty$ of fixed energy $E$, whose distribution is given by the bath that interests us. The asymptotic incoming flux is correctly normalized, despite the fact that $e^{-\text{Im}(\chi)}$ is a real amplitude modulation near the scattering region. This NLO correction, which corresponds to different normalizations for the various channels, is the source of non-zero energy transfer.

We can now define the many-body bath operator for the $P$ particles, whose correlators will determine the response of the spin system fluctuations. We have

\begin{equation}
    \Psi(\mathbf{x},t)=\sum_\mathbf{k} c_\mathbf{k} \psi_\mathbf{k}(\mathbf{x},t),
    \label{eq:manybody}
\end{equation}
with $\avg{c_{\mathbf{k}'}^\dagger c_\mathbf{k}}=f(\mathbf{k})\delta_{\mathbf{k},\mathbf{k}'}$ in an incoherent bath, where $f(\mathbf{k})$ is the incoming, asymptotic, energy distribution of the bath particles, and $\psi_\mathbf{k}=\psi_E$ for $E=k^2/(2m)$ in the reference frame whose $z$-axis is aligned with $\mathbf{k}$.

\subsection{Spin response}
\label{app:spin_response}

The Holstein--Primakoff bosons that describe the fluctuations of the spin system evolve according to the Hamiltonian derived in the main text,

\begin{equation}
 H_{I,i}
\simeq \int\di^3
\mathbf{r}\,
g_i(\mathbf r)
\left\{
2\,\operatorname{Im}\!\left[
\mathcal O_{I,+}e^{-i\os t}
\right]s_y^{(i)} + 
\mathcal O_{I,z}s_z^{(i)}
\right\}.
\label{eq:HI_spin}
\end{equation}

Take $\mathcal O_{I,+}=\Psi^\dagger \Gamma_+ \Psi$, where $\Gamma_+$ is an internal-space operator, and $\Psi$ the many-body operator defined in Eq.~\eqref{eq:manybody}. The total density matrix of the spin system obeys a Lindblad equation once the bath is traced out. The mean of Eq.~\eqref{eq:HI_spin}, $H_\text{eff}=\avg{H_I}_B=\text{tr}[\rho_B H_I]$, where $\rho_B$ is the density matrix of the bath, gives unitary evolution, while the fluctuations lead to decoherence.  Taking

\begin{equation}
    H_\text{eff} = 2\sum_i\int\di^3\mathbf{r}\, \left\{g_i(\mathbf{r}) \text{Im}[\avg{\mathcal{O}_{I,+}}_B e^{-i\os t}]s_y^{(i)} + \avg{\mathcal{O}_{I,z}}_B s_z^{(i)}\right\},
\end{equation}
and the correlation functions

\begin{align}
    C_{y}^{ij}(\tau,t)&=4\int\di^3\mathbf{r}\,\di^3\mathbf{r}'\, g_i(\mathbf{r})g_j(\mathbf{r}') \avg{\text{Im}[\mathcal{O}(\mathbf{r},t)e^{-i\os t}]\text{Im}[\mathcal{O}(\mathbf{r}',t')e^{-i\os t'}]}_B\\
    C_z^{ij}(\tau,t)&=\int\di^3\mathbf{r}\,\di^3\mathbf{r}'\, g_i(\mathbf{r})g_j(\mathbf{r}') \avg{\mathcal{O}_{z}(t)\mathcal{O}_{z}(t')}_B,
\end{align}
where $\tau=t-t'$, we find the master equation for the system fluctuations $\rho_S$

\begin{equation}
    \dot{\rho}_S = i[\rho_S,H_\text{eff}] + \sum_{ij}\gamma_y^{ij}\parea{
s_y^{(j)}\rho_S s_y^{(i)}
-\frac12
\left\{
s_y^{(i)}s_y^{(j)},\rho_S
\right\}} + \sum_{ij}\gamma_z^{ij}\parea{
s_z^{(j)}\rho_S s_z^{(i)}
-\frac12
\left\{
s_z^{(i)}s_z^{(j)},\rho_S
\right\}},
\end{equation}
where the rates are

\begin{align}
    \gamma_y^{ij} &= 2\int_{0}^{t} \di \tau\, \bar{C}_y^{ij}(\tau),\\
    \gamma_z^{ij} &= 2\int_{0}^{t} \di \tau\, \bar{C}_z^{ij}(\tau),
    \label{eq:master}
\end{align}
where $\bar{C}$ is $C$ averaged over a long $T\gg\os^{-1}$ to remove secular terms, and we assumed zero $yz$ cross-correlation.

We emphasize that the master equation in this form, Eq.~\eqref{eq:master}, applies to the Holstein--Primakoff bosons that describe fluctuations around the equatorial coherent spin state. It does \emph{not} apply, for instance, to a spin system in its physical ground state.

We will focus on $\gamma_y^{ij}$ in what follows, since this term induces energy spread. The $\gamma_z^{ij}$ commutes with $J_z$, and so can only induce dephasing.

Expectation values can be expanded using the many-body operators Eq.~\eqref{eq:manybody},
\begin{equation}
\begin{split}
\avg{\mathcal{O}_\alpha}_B &= \avg{{\Psi}^\dagger\Gamma_\alpha\Psi}_B\\
&= \sum_{\mathbf{k},\mathbf{k}',s,s'}\avg{c_{\mathbf{k},s}^\dagger c_{\mathbf{k}',s'}}_B \psi_{\mathbf{k},s}^\dagger\Gamma_\alpha\psi_{\mathbf{k}',s'}\\
&=\sum_{\mathbf{k},s}f_s(\mathbf{k}) \psi_{\mathbf{k},s}^\dagger\Gamma_\alpha\psi_{\mathbf{k},s},
\end{split}
\end{equation}
where $\mathbf{k}$ denotes the momentum and $s$ the internal degrees of freedom of $P$, $f_s(\mathbf{k})$ is the momentum distribution per internal degree of freedom in an incoherent bath, and $\Gamma_\alpha$ are internal space operators acting on the wavefunctions $\psi_{\mathbf{k},s}$, such as particle momentum or spin.

Similarly

\begin{equation}
\begin{split}
\avg{\mathcal{O}_\alpha(t)\mathcal{O}_\beta(t')}_B =\sum_{\mathbf{k},s,\mathbf{k}',s'}f_s(\mathbf{k}) \parea{1\pm f_{s'}(\mathbf{k'}) }\psi_{\mathbf{k},s}^\dagger(t)\Gamma_\alpha\psi_{\mathbf{k}',s'}(t)\psi_{\mathbf{k}',s'}^\dagger(t')\Gamma_\alpha\psi_{\mathbf{k},s}(t'),
\end{split}
\end{equation}
where we dropped terms that are independent of $\tau=t-t'$, and $\pm f_{s'}(\mathbf{k}')$ corresponds to scattering of bosons and fermions, respectively. Because Pauli blocking (or Bose enhancement) will generally not concern us in this work, we will be dropping $f(\mathbf{k}')$.

\subsection{Observables}
\label{app:observables}

Using the master Eq.~\eqref{eq:master} we can now compute the rates for energy transfer and decoherence. Mean energy-transfer arises from $H_\text{eff}$, while energy decoherence from $\gamma_y$. We have

\begin{align}
    \frac{\di \avg{J_z}}{\di t} &= -\sum_i\int\di^3 r g_i(\mathbf{r}) \text{Im}[\avg{\mathcal{O}_{I,+}}_B e^{-i\os t}]\\
    \frac{\di \avg{J_z^2}}{\di t} &=\frac{1}{4}\sum_{ij}\gamma_y^{ij}.
\label{eq:obs_fl}
\end{align}

To compute energy exchange we use $V_+e^{-i
\os t} = \pare{V_- e^{i
\os t}}^\dagger=\frac{1}{2}\sum_i g_i(\mathbf{r})\Gamma_+ e^{i\os t}$, $V^{-1}\sum_k\to \int\frac{\di^3 k}{(2\pi)^3}$ and $\psi_k^\dagger \psi_k = e^{-2\text{Im}(\chi)}$ within the eikonal approximation, where if $\Gamma$ is momentum-dependent, then it is evaluated at momentum $\mathbf k$. This yields directly Eq.~\eqref{eq:jz_gen}. Subsequent substitution of the NLO eikonal phase, Eq.~\eqref{eq:imag_eik}, yields Eq.~\eqref{eq:jz} after direct algebraic manipulation and time-averaging to remove secular terms oscillating at frequency $\os$.

In the eikonal approximation one can compute the decoherence rate directly. We have the differential eikonal phase in the expectation value of bath operators, which is $\Delta\chi=\chi_{k}(t,\mathbf{r})-\chi_{k'}(t,\mathbf{r})$. This can be approximated in the eikonal limit $\frac{\Delta k}{k}\approx \frac{(n\pm1)\os}{kv}\ll1$ as

\begin{equation}
    \Delta\chi \approx\frac{(n\pm1) q_0}{k}\frac{\partial}{\partial q_0}\pare{q_0\chi_k}\equiv (n\pm1)g_k(\theta),
\end{equation}
$\theta=\omega t$. The function $g_k$ is of order $\frac{q_0}{k}\chi\sim \frac{V}{kv}\ll 1$, so that $e^{ig_k}\approx 1$, but in principle $ng_k$ can be large. Nevertheless, when time-averaged against $e^{-in\os t}$ it gives

\begin{equation}
    I_n=\int_0^{2\pi}\frac{\di\theta}{2\pi}e^{in\parea{g_k(\theta)-\theta}},
\end{equation}
with the integrand periodic in $\theta$. As a result, this is non-zero only if there exist stationary points of $g_k(\theta)-\theta$, but $|g'_k|\sim \mathcal{O}(g_k)$, and so the stationarity condition cannot be satisfied $g'_k(\theta_0)=1$, leading to $I_n\sim g_k^n$. Thus only the $n=0$ term survives to leading order, yielding eventually

\begin{equation}
    \frac{\di\avg{J_z^2}}{\di t} \approx \int\frac{\di^3 k}{(2\pi)^3}\frac{\di^3 k'}{(2\pi)^3} f(\mathbf{k})\parea{1\pm f(\mathbf{k}')} 2\pi\parea{\delta(E-E'-\os)\abs{\widetilde{V}(\mathbf{q})}^2+\delta(E-E'+\os)\abs{\widetilde{V}(\mathbf{q})}^2},
\end{equation}
equal to the Born approximation result. This can also be proved by expanding the $e^{ing_k}$ into Bessel functions, in which case it can be shown that, parametrically, the time-average picks $J_n(n \epsilon)$, with $\epsilon\ll1$ being order $g_k$. By the Debye approximation for $\epsilon\ll1$, $J_{|n|}(|n|\epsilon)\approx (2\pi |n|)^{-1/2} (\text{e} \epsilon/2)^{|n|} e^{-|n|\epsilon^2/4}$, and again the leading term is $n=0$.

One can use Floquet theory to derive a relation between the decoherence rate and the cross-sections to exchange $n$ quanta. We refer the reader who is unfamiliar with Floquet theory to the next section and the references therein.

First, we make the observation that $\sum_i \frac{1}{2}g_i(\mathbf{r})\text{Im}[\Gamma_+e^{-i\os t}]=\frac{1}{\os}\partial_t V(\mathbf{r},t)$, and so

\begin{equation}
    \frac{1}{4}\sum_{ij}\gamma_y^{ij} = \frac{1}{\os^2}\int\frac{\di^3 k}{(2\pi)^3}\frac{\di^3 k'}{(2\pi)^3} f(\mathbf{k})\parea{1\pm f(\mathbf{k}')} 2\pi\sum_n \delta(E-E'+n\os)|c_n|^2,
\end{equation}
where

\begin{equation}
    c_n = \int\di^3 r\,\int_0^{2\pi/\os}\frac{\di t}{2\pi/\os} \pare{u_k^{(+)}}^\dagger \dot{V}u_{k'}^{(+)}e^{-in\os t}.
\end{equation}
where we isolated the energy $\psi_k=e^{-iEt}u_k^{(+)}(\mathbf{r},t)$, and so the $u_k^{(+)}$ are periodic functions. We are now including the superscript $+$ because in the preceding section we computed the incoming scattering states, which form a complete basis and are the states that we expanded the many-body bath operator in. In Floquet theory, one defines the dot product as the average over a period,
\begin{equation}
\langle\!\langle u|v\rangle\!\rangle
=
\frac1{2\pi/\os}\int_0^{2\pi/\os}dt\,
\langle u(t)|v(t)\rangle
=
\frac1{2\pi/\os}\int_0^{2\pi/\os}dt
\int d^3r\,u^\dagger(\mathbf r,t)v(\mathbf r,t),
\end{equation}
and so $c_n = \langle\!\langle u_k^{(+)}|\dot{V}|u_{k',n}^{(+)}\rangle\!\rangle$, with $u_{k',n}^{(\pm)} = u_{k'}^{(\pm)} e^{-in
\os t}$. In the space of $|u\rangle\!\rangle$, called the Sambe space, one can show that completeness of the scattering basis $u_k^{(+)}$ implies 

\begin{equation}
    \sum_n\int\frac{\di^3 k}{(2\pi)^3}|u^{(+)}_{k,n}\rangle\!\rangle\langle\!\langle u^{(+)}_{k,n}| = \sum_n\int\frac{\di^3 k}{(2\pi)^3}|u^{(-)}_{k,n}\rangle\!\rangle\langle\!\langle u^{(-)}_{k,n}|,
\end{equation}
where the superscript $-$ denotes scattering states arriving from $+\infty$. We assume there are no bound states here. Furthermore, $u_{k,n}^{(\pm)}$ is an eigenstate of $\mathcal{K}=H(t)-i\partial_t$ with eigenvalue $E_k-n\os$, and so $\delta(E-E'+n\os)u_{k',n}^{(\pm)}=\delta(E-\mathcal{K})u_{k',n}^{(\pm)}$. 

Dropping the final state occupation $f(\mathbf{k}')$ (assuming no Pauli blocking or Bose enhancement) for a bath away from degeneracy, we can use the completeness relation to write

\begin{equation}
        \frac{1}{4}\sum_{ij}\gamma_y^{ij} = \frac{1}{\os^2}\int\frac{\di^3 k}{(2\pi)^3}\frac{\di^3 k'}{(2\pi)^3} f(\mathbf{k}) 2\pi\sum_n \delta(E-E'+n\os)\langle\!\langle u_k^{(+)}|\dot{V}|u_{k',n}^{(-)}\rangle\!\rangle\langle\!\langle u_{k',n}^{(-)}|\dot{V}|u_k^{(+)}\rangle\!\rangle.
\end{equation}

Now we define the class of potentials $V_\phi$ by taking $\os t\to \os t+\phi$ in $V$. Then $\os^{-1}\dot{V}=\partial_\phi V$. The scattering Hellmann-Feynman identity gives for the transfer matrix $\partial_\phi T_{k',k}^{(n)}(\phi)=\langle\!\langle u_{k',n}^{(-)}|\partial_\phi V|u_k^{(+)}\rangle\!\rangle$. By periodicity, $T_{k',k}^{(n)}(\phi)=e^{-in\phi}T_{k',k}^{(n)}(0)$, and so inside the channel sum and final phase space integral, we can substitute $|c_n|^2$ by $\os^2n^2\abs{T_{k',k}^{(n)}(0)}^2$. The square of the transfer matrix is directly proportional to the channel differential cross-section, as defined in the next section and the main text. The sum over final phase-space then gives

\begin{equation}
    \frac{\di\avg{J_z^2}}{\di t} = \int\frac{\di^3 k}{(2\pi)^3}f(\mathbf{k})v\sum_{n=-\infty}^{+\infty} n^2\sigma_n. 
    \label{eq:app:Jz2_distribution}
\end{equation}

We provide a third proof of this relation in the next section, from $P$'s perspective.

\subsection{Floquet-eikonal scattering}
\label{app:Floquet}

Herein we develop the application of the eikonal approximation to a time-periodic sinusoidal potential in the context of quantum mechanics. Glauber treated general time dependence in eikonal scattering theory \cite{Glauber:eikonal}. Recently, Ref.~\cite{Liu2024} connected the eikonal approximation with Floquet potentials \cite{Floquet1883, Shirley1965, Zeldovich1967, Sambe1973} and exploited the optical theorem to compute total cross sections. For a \textit{sinusoidal} potential in time, we derive the Floquet channel structure in terms of $\sigma_n$ and compute their first and second moments.

The time-periodicity of the potential compels a structure of energy conservation in the $S$-matrix where only discrete quanta of energy $\omega_0$ can be exchanged \cite{Howland1974, Yajima1977, Howland1979, Nakamura1986, Floquet:review}. 
\begin{equation}
    S_{\mathbf{k}'\mathbf{k}} = \delta^3(\mathbf{k}'-\mathbf{k}) + \frac{i}{2\pi m}\sum_{n=-\infty}^\infty \delta(E_\mathbf{k}-E_{\mathbf{k}'}+n\omega_0) f_{\mathbf{k}'\mathbf{k}}
\end{equation}

Under the standard eikonal conditions of $V\ll E$ and small deflection/diffraction in the central region of interaction, along with $|n|\omega_0\ll E$ for the channels that contribute significantly, the change of variable $\tau\equiv t-z/v$ \cite{Glauber:eikonal} yields
\begin{align} \label{eq:eikonal:amplitude}
 f_{\mathbf{k}'\mathbf{k}} \approx \frac{-ik\omega_0}{(2\pi)^2}\int d^2\mathbf{b}\ e^{i\mathbf{q}_\perp\cdot\mathbf{b}} \int_0^{2\pi/\omega_0} d\tau\ e^{in\omega_0 \tau} \left(e^{i\chi(\tau,\mathbf{b})}-1\right).
\end{align} The eikonal phase here is the line integral along the classical trajectory:
\begin{align}
\chi(\tau,\mathbf{b})&\equiv \frac{-1}{v}\int_{-\infty}^\infty V\left(\tau+\frac{z}{v},\mathbf{b},z\right)dz.
\end{align}

As argued in the letter, the interaction with the macroscopic target can be modeled with a semiclassical Hermitian potential:
\begin{equation} \label{eq:potential:harmonic}
    V(t,\mathbf{r})\equiv V_-(\mathbf{r})e^{i\omega_0 t} + V_+(\mathbf{r})e^{-i\omega_0 t} + V_0(\mathbf{r}).
\end{equation}
For generality, we added a Hermitian time-independent component $V_0$. For such a potential, the eikonal phase can be put in the form:
\begin{equation}
    \chi(\tau,\mathbf{b}) = -\sin\left[\omega_0\tau-\delta(\mathbf{b})\right] \bar{\chi}(\mathbf{b})+\chi_0(\mathbf{b})
\end{equation}
where $\chi_0$ is due to $V_0$, and $\bar\chi$ to the inelastic part of the potential. Then, the Jacobi-Anger identity can be applied to perform the temporal integral:
\begin{equation}
    f_{\mathbf{k}'\mathbf{k}} = \frac{-ik}{2\pi} \int d^2\mathbf{b}\ e^{i\mathbf{q}_\perp\cdot\mathbf{b}} \left(J_n[\bar\chi(\mathbf{b})]e^{i\chi_0(\mathbf{b})+i n \delta(\mathbf{b})}-\delta_{0n}\right).
\end{equation}

The cross sections for various Floquet channels are organized as
\begin{equation} \label{eq:app:sigman}
\sigma_n = \int d^2\mathbf{b} \left|J_n[\bar\chi(\mathbf{b})]e^{i\chi_0(b)}-\delta_{0n}\right|^2.
\end{equation}
The total cross section obtained from the optical theorem $\sigma_{\rm tot} = \frac{4\pi}{k} {\rm Im} f_{\mathbf{k}\mathbf{k}}$ agrees with $\sum_n \sigma_n$, though the eikonal approximation itself fails for $|n|\omega_0\sim E$:
\begin{equation}
    \sigma_{\rm tot} = 2\int d^2\mathbf{b}\left(1-\cos\chi_0(\mathbf{b})J_0[\bar\chi(\mathbf{b})]\right).
\end{equation}

The second moment of the Floquet cross sections $\sum_n n^2\sigma_n$ is related to the target's decoherence and can be directly computed from $\sigma_n$'s given above, which yields \eqref{eq:jz2_born_identity}. Given their relation to the change in a spin ensemble's variance (\ref{eq:app:Jz2_sigman}), demonstrated below, we reproduce (\ref{eq:jz2}).

At LO in the eikonal conditions, $\sigma_n=\sigma_{-n}$, so the net rate of energy transfer ($\propto -\sum_n n\sigma_n$) vanishes. A NLO computation from the scattering picture is undertaken in \ref{app:NLO_eikonal}. The result agrees with (\ref{eq:jz}) given the identification $\dot{E}_{P}=-\omega_0\dot{J}_z$, though this derivation of $\dot{E}_P$ from the scattering picture is agnostic about the nature of the target.

\subsection{Spin diffusion from the $S$-matrix}
\label{app:decoherence_S}

The scattering particle $P$ and the target $\psi$ form a coupled system. 
Our target system is an ensemble of spins:
$ H_\psi=\omega_0 J_z\equiv \omega_0 \sum_i S_z^{(i)}$. Consider projections of the full $S$-operator to the subspace of $\psi$: specifying $P$'s state by its momentum,
\begin{equation}
M_n(\mathbf{p}',\mathbf{p})\equiv \langle\mathbf{p}'| S |\mathbf{p}\rangle
\end{equation}
are Kraus operators \cite{Kraus,Janzing:2004nzg,Hornberger:2008xkz} on $\psi$. We further \emph{restrict} the energy transfer $E_{\mathbf{p}'}-E_\mathbf{p} = n\omega_0$ to integer $n$. Then, from $[S,H_P+H_\psi]=0$, involving the free Hamiltonians 
\cite{Taylor:1972pty,Janzing:2004nzg}, we have the commutation 
\begin{equation} \label{eq:app:nsq:commute}
    [J_z,M_n] = -n M_n
\end{equation}
that shows $M_n$ is a lowering operator on the target's $J_z$ by $n$ units.

For a given initial product state $|P,\psi\rangle \equiv |\mathbf{p}\rangle|\psi\rangle$, consider a projection operator $\mathcal{P}_n^P$ in $P$'s subspace:
\begin{equation}
    \mathcal{P}_n^P = \int d\Pi_{n}^{p'} |\mathbf{p}'\rangle\langle\mathbf{p}'|
\end{equation}
where $\int d\Pi_{n}^{p'}$ integrates over the phase space of $n$ quanta of energy exchange. The corresponding scattering probability is, for $n \neq 0$,
\begin{align}
    P_n  &\equiv n_\chi\sigma_n v\Delta t =\langle\mathbf{p},\psi|S^\dagger (\mathcal{P}_n^P \otimes \mathbf{1}_\psi) S |\mathbf{p},\psi\rangle \nonumber\\
    &= \langle\psi|\mkern-5mu\int d\Pi_{n}^{p'} M_n^\dagger(\mathbf{p},\mathbf{p}') M_n(\mathbf{p}',\mathbf{p})|\psi\rangle 
\end{align}
where $\Delta t$ is a formal scattering time. From $S^\dagger S=1$ and $\sum_n \mathcal{P}_n^P=\mathbf{1}_P$, we also have
\begin{equation} \label{eq:app:nsq:completeness}
    \sum_n \int d\Pi_n^{p'} M_n^\dagger(\mathbf{p},\mathbf{p}') M_n(\mathbf{p}',\mathbf{p}) = \mathbf{1}_\psi. 
\end{equation}

By interaction, $|\mathbf{p},\psi\rangle\to S|\mathbf{p},\psi\rangle$, so
\begin{align}
    &\Delta\langle J_z^2\rangle \equiv \langle\mathbf{p},\psi|S^\dagger J_z^2 S|\mathbf{p},\psi\rangle - \langle\mathbf{p},\psi|J_z^2|\mathbf{p},\psi\rangle \nonumber\\
    =& \sum_n n^2 P_n - 2\sum_n n \langle\psi|\mkern-5mu \int d\Pi_n^{p'} M_n^\dagger M_n J_z |\psi\rangle. \label{eq:app:decoherence:two_terms}
\end{align}
The completeness relation (\ref{eq:app:nsq:completeness}) and the commutation (\ref{eq:app:nsq:commute}) were employed to obtain this result. In the second term in (\ref{eq:app:decoherence:two_terms}), $J_z|\psi\rangle$ yields, for our equatorial product state, a superposition of states, each of which has one spin modified. For a macroscopic target, this is expected to cause a negligible backreaction on the scattering rates. Thus $\sum_n n\sigma_n\approx 0$ at LO entails that this second term is subdominant. Expressing the rate of interaction with the cross sections, we obtain:
\begin{equation} \label{eq:app:Jz2_sigman}
    \frac{d\langle J_z^2\rangle}{dt}\approx n_\chi v \sum_n n^2 \sigma_n.
\end{equation}
Appending a distribution over the initial state $|\mathbf{p}\rangle$ and employing (\ref{eq:jz2_born_identity}), this validates \eqref{eq:app:Jz2_distribution} stated earlier.

\subsection{NLO eikonal expansion for energy transfer} \label{app:NLO_eikonal}

Under the (LO) eikonal approximation, the channel cross sections reported in the letter do not discriminate between $\sigma_{\pm n}$, yet certain observables depend precisely on their difference. To compute the rate of energy transfer $\propto\frac{d}{dt}\langle J_z\rangle\propto -\sum_n n\sigma_n$, one must proceed to higher orders in the eikonal conditions. Although there are more expedient methods of deriving this, as given in the letter, here we perform a systematic expansion of the corrections at NLO to illustrate the scattering formalism \cite{Wallace1973}. The eikonal approximation can be viewed as a perturbative expansion, classically speaking, in $V/E$, the deflection angle $\theta$, and the inelasticity $\varepsilon\equiv \omega_0/E\propto \qmin/k$, for $\qmin\equiv\omega_0/v$. Here we use $E\equiv k^2/2m$ for the (asymptotic) kinetic energy. For \textit{notational brevity}, we take $\omega\equiv\omega_0$ in this supplemental section. As we'll see, to the first moment of channel cross sections, the latter two contribute. By formulating the physics in channel space, it shows the origins of the NLO corrections, many of which cancel in a systematic fashion.

In this section, we consider a Hermitian harmonic potential of form
\begin{equation}
    V(t) = V_- e^{i\omega_0 t} + V_+ e^{-i\omega_0 t} + V_0 \equiv U(t)+V_0
\end{equation}
where $V_+=V_-^\dagger$. For convenience, we also define the complex phase
\begin{equation}
    \zeta(\mathbf{b}) = \int_{-\infty}^\infty e^{i\qmin z} V_-(z)\ dz
\end{equation}
that is related to the eikonal phase as
\begin{equation} \label{eq:eikonal:phase:complex_parts}
    \chi(\tau) = \frac{-1}{v}\int_{-\infty}^\infty V\left(\tau+\frac{z}{v},z\right) dz = \frac{-1}{v}\left(e^{i\omega_0\tau}\zeta + e^{-i\omega_0\tau}\zeta^* \right)+\chi_0.
\end{equation}
When these phase functions are not integrated up to $z=\infty$ but up to some finite $z$, as appears in the wave function, we place a $\sim$ on top of them, e.g. $\tilde{\chi}(\tau,z)$.

\subsubsection{First moment} \label{app:observable:form}

Schematically, the amplitude can be expressed as
\begin{equation}
    f_{\mathbf{k}'\mathbf{k}} = \frac{-ik\omega_0}{(2\pi)^2}\int d^2\mathbf{b}\ e^{i\mathbf{q}_\perp\cdot\mathbf{b}}\int_0^{2\pi/\omega_0} d\tau\ e^{in\omega_0\tau} F(\tau,\mathbf{b})
\end{equation}
where
\begin{equation}
    F(\tau,\mathbf{b})\equiv \frac{-i}{v}\int_{-\infty}^\infty e^{i(q_z+n\qmin)z} V\varphi\  dz
\end{equation}
is taken to be \textit{independent of} $n$. We shall demonstrate this is so at NLO.

Perturbatively, with $F\approx F^{(0)} + F^{(1)}$, it follows
\begin{equation} \label{eq:NLO:observable}
    -\sum_n n\sigma_n \approx \frac{\varepsilon}{v^2}\int d^2\mathbf{b} |\zeta|^2+\frac{1}{\pi}\int d^2\mathbf{b} \int_0^{2\pi/\omega_0} \partial_\tau\chi\ {\rm Re}\left(e^{-i\chi}F^{(1)}\right) d\tau.
\end{equation}
The first term comes from phase-space corrections of flux and $d\Omega=\frac{d^2\mathbf{q}_\perp}{k_n^2\cos\theta}$ when integrating to obtain the cross section; the second issues from the corrections in the scattering amplitude. In passing, we note that for the phase-space term, the effect from $\cos\theta\neq 1$ is subleading compared to the diffractive corrections we consider subsequently. As is made clear later, the amplitude corrections $F^{(1)}$ naturally contains the phase $e^{i\chi}$, due to the LO wave function being a pure phase, so selecting the real part removes a number of corrections which would otherwise enter for the wave function and amplitude themselves.

\subsubsection{Wave-function corrections}
\label{app:observable:wave-function}

\newcommand{\cwf}[2]{\ensuremath{\varphi_{#1}^{(#2)}}}
\newcommand{\gwf}[1]{\ensuremath{\Phi^{(#1)}}}

To construe the physical origin of the relevant corrections, it is useful to begin with the decomposition of the Schr\"{o}dinger equation to Floquet channels. The organization of NLO perturbations is rendered transparent via this method. In this section, we assume a potential $V(\mathbf{r})$, which could describe an interaction independent of the spin of the scattering particle's spin and momentum. The millicharge scenario discussed in the paper, nonetheless, has a momentum-dependent potential, and conceivably $\hat{\mathbf{p}}$ could induce NLO corrections. Thus, in the final subsection, we show that for electromagnetic couplings expressed in the gauge $\nabla\cdot\mathbf{A}=0$, the additional contribution from $\mathbf{A}\cdot\hat{\mathbf{p}}$ acting on the envelope vanishes for energy transfer. There, we also circumvent the decomposition into Floquet channels.

We adopt the ansatz, in accord with the time-periodicity of the Floquet wave function $\phi$ \cite{Shirley1965, Sambe1973, Howland1979},
\begin{equation}
    \psi(t,\mathbf{r}) \equiv e^{-iEt}\sum_{n=-\infty}^\infty e^{-in\omega_0 t} \frac{e^{i k_n z}}{(2\pi)^{3/2}}\varphi_n(\mathbf{r}) \equiv \frac{e^{-iEt+ikz}}{(2\pi)^{3/2}}\varphi(t,\mathbf{r})
\end{equation}
where $k_n^2\equiv k^2+2mn\omega_0$. In the time-independent picture of scattering, the $n^{\rm th}$ Floquet amplitude is given by the asymptotic behavior of $\varphi_n$ \cite{Liu2024}. Under the above ansatz, the \textit{channel wave equations} become
\begin{equation} \label{eq:NLO:channel_eq}
    \frac{i k_n}{m} \partial_z\varphi_n + \frac{\nabla^2}{2m}\varphi_n = V_- e^{i(k_{n+1}-k_n)z}\varphi_{n+1} + V_+ e^{i(k_{n-1}-k_n)z}\varphi_{n-1} +V_0\varphi_n.
\end{equation}

In the LO eikonal approximation, the $\nabla^2$ term is dropped. Furthermore, we take the ``channel velocity factor" $k_n/m\approx k/m=v$ and approximate the phase difference with $k_{n}-k_{n-1}\approx \qmin$:
\begin{equation} \label{eq:NLO:LO_channel_eq}
    i v\partial_z\varphi_n^{(0)} = V_- e^{i\qmin z}\varphi_{n+1}^{(0)} + V_+ e^{-i\qmin z} \varphi_{n-1}^{(0)}+V_0\cwf{n}{0}.
\end{equation}

This can readily be solved by introducing a generating function/Fourier series:
\begin{equation}
    \Phi^{(0)} (\tau) \equiv \sum_{n=-\infty}^\infty e^{-in\omega_0\tau}\cwf{n}{0}
\end{equation}
where $\tau$ here is a \emph{free parameter} in the time-independent channel space, though shortly we'll see that choosing it as $t-z/v$ produces the LO eikonal wave function. The LO channel wave equation transforms to an equation for $\gwf{0}$:
\begin{equation}
    iv\partial_z\gwf{0} = V\left(\tau+\frac{z}{v}\right) \gwf{0}.
\end{equation}
This is the LO eikonal wave equation, and $\gwf{0}$ is the LO time-dependent eikonal wave function under the \textit{identification} $\tau=t-z/v$. Had one dropped the phases $e^{\pm i\qmin z}$ in (\ref{eq:NLO:LO_channel_eq}), the temporal retardation of the potential would be omitted.

At NLO, in (\ref{eq:NLO:channel_eq}) both the velocity factor $v_n\approx v(1+\frac{n}{2}\epsilon)$ on the LHS and the phases on the RHS $\exp[i(k_{n+1}-k_n)z-i\qmin z]\approx\exp[-i(2n+1)\qmin z\varepsilon/4]$ receive $\mathcal{O}(\varepsilon)$ corrections:
\begin{align}
    &iv\partial_z\cwf{n}{1} - \left(V_- e^{i\qmin z}\cwf{n+1}{1} + V_+ e^{-i\qmin z}\cwf{n-1}{1}+V_0\cwf{n}{1}\right) = C_{v_n} + C_\phi + C_{\nabla^2}
\end{align}
where the source terms $C_{v_n}, C_\phi, C_{\nabla^2}$ are the following functions:
\begin{align} \label{eq:NLO:NLO_channel_eq}
    C_{v_n} &\equiv \frac{-inv\varepsilon}{2}\partial_z\cwf{n}{0} \\
    C_\phi &\equiv \left(e^{-i\frac{2n+1}{4}\varepsilon\qmin z}-1\right) V_-e^{i\qmin z}\cwf{n+1}{0} + \left(e^{i\frac{2n-1}{4}\varepsilon\qmin z}-1\right)V_+ e^{-i\qmin z}\cwf{n-1}{0} \\
    C_{\nabla^2} &\equiv -\frac{\nabla^2}{2m}\cwf{n}{0}.
\end{align}
The LHS of the NLO channel equation (\ref{eq:NLO:NLO_channel_eq}) has the same form as (\ref{eq:NLO:LO_channel_eq}), and it can be analogously solved with a generating function
\begin{equation}
    \gwf{1}(\tau) \equiv \sum_{n=-\infty}^\infty e^{-in\omega_0\tau}\cwf{n}{1}
\end{equation}
where again $\tau$ is a free parameter at this stage. The NLO channel wave equation is transformed to
\begin{equation} \label{eq:NLO:sources}
    \left[iv\partial_z - V\left(\tau+\frac{z}{v}\right)\right]\gwf{1} = G_{v_n} + G_\phi + G_{\nabla^2}\equiv \mathcal{D}\gwf{0}
\end{equation}
in which $G_{v_n}, G_\phi, G_{\nabla^2}$ are the source terms corresponding to $C_{v_n}, C_\phi, C_{\nabla^2}$. Their content will be elaborated shortly; for now, we merely observe that at NLO they all naturally contain a factor of $\gwf{0}$ and can also be viewed as the result of a differential operator acting on $\gwf{0}$. This $1^{\rm st}$ order DE can be solved with the integrating factor $\left(\gwf{0}\right)^*$:
\begin{equation} \label{eq:NLO:gwf}
    \gwf{1}(z) = \frac{-i}{v}\gwf{0}(z)\int_{-\infty}^z  \left(\gwf{0}(z')\right)^*\mathcal{D}\gwf{0}(z') dz'.
\end{equation}
Having solved the NLO channel equations, we're in a position to construct the NLO wave function.
\begin{align}
    \varphi(t,\mathbf{r})
    &= \sum_n e^{-in\omega_0 (t-z/v)} e^{i(k_n-k-n\qmin)z}\varphi_n \nonumber \\
    &= e^{i(\hat{K}_c-k-i\partial_\tau/v)z}\Phi(\tau,\mathbf{r}) \equiv e^{\hat\alpha z}\Phi
\end{align}
where we've chosen to evaluate the generating function $\Phi(\tau)\equiv\sum_n e^{-in\omega_0\tau}\varphi_n$ at $\tau\equiv t-z/v$ to express the time-dependent wave function. Time-differentiation operators are defined to reproduce the appropriate phase factors when acting on the Floquet channels: $K_c^2\equiv k^2+2m i\partial_\tau$ is the operator for $k_n^2$. From differentiation, it follows
\begin{equation}
    e^{-\alpha z} \partial_z e^{\alpha z} = \alpha+\partial_z.
\end{equation}
This approach factors out the free propagation of the wave, akin to operator definitions in the interaction picture, which facilitates a subsequent cancellation in the amplitude \cite{Yahiro2011}.

In perturbation theory, we shall organize corrections that approximate $\Phi$. Thus we return to the source terms for $\gwf{1}$. They assume the following form:
\begin{align}
    G_{v_n} &\equiv \frac{\varepsilon}{2\qmin}\partial_\tau\partial_z\gwf{0} \label{eq:NLO:sources:vn}\\
    G_\phi &\equiv  V_-e^{i(\qmin z+\omega_0\tau)}\left[e^{\frac{i}{4}\varepsilon\qmin z}\gwf{0}\left(\tau+\frac{\qmin z}{2 E}\right)-\gwf{0}(\tau)\right] \nonumber\\
    &\quad\quad\quad\quad + V_+ e^{-i(\qmin z+\omega_0\tau)} \left[e^{\frac{i}{4}\varepsilon\qmin z}\gwf{0}\left(\tau-\frac{\qmin z}{2 E}\right)-\gwf{0}(\tau)\right] \label{eq:NLO:sources:phi}\\
    &\approx \frac{\varepsilon}{4}\qmin z\gwf{0}\left[iU\left(\tau+\frac{z}{v}\right) + {\rm Re}\#\right]\\
    G_{\nabla^2} &\equiv -\frac{\nabla^2}{2m}\gwf{0} \label{eq:NLO:sources:laplacian}.
\end{align}
We took $|n|\varepsilon\qmin z\ll 1$ to linearize the NLO exponential phase factors. Although this is certainly not justified at very large $z$, we assume it is protected by the suitable decay of the potential's envelope. Also, only the imaginary part is retained in the square brackets, which contributes to the observable, as shown later. Indeed, among these three source terms which correct the wave function, only $\gwf{1}_{\phi}$ ultimately enters nontrivially into the observable.

\subsubsection{Amplitude and first-moment corrections}
The corrections to the wave function enter the amplitude via
\begin{equation} \label{eq:NLO:Schro}
    V\varphi = i v\partial_z\varphi + \frac{1}{2m}\left(\partial_z-\frac{1}{v}\partial_\tau\right)^2\varphi + \frac{1}{2m}\nabla_\perp^2\varphi
\end{equation}
where the longitudinal Laplacian is complicated due to the temporal change of variable. Now, the spatial derivatives are defined at fixed $\tau$. Notably, this is no longer an exact derivative, as in the LO eikonal case, which could be trivially integrated. With $\varphi=e^{\alpha z}\Phi$, the longitudinal part of the amplitude can be expressed as
\begin{align}
\int_0^{2\pi/\omega_0}d\tau\,e^{in\omega_0\tau} F(\tau,\mathbf{b}) &= \frac{-i}{v} \int_0^{2\pi/\omega_0}d\tau\,e^{in\omega_0\tau}\int dz\, e^{i(k_n-k_z')z}e^{-i(k_n-k-nq_0)z} V e^{i(K_c-k-i\partial_\tau/v)z} \Phi \nonumber\\
    &= \frac{-i}{v} \int_0^{2\pi/\omega_0}d\tau\,e^{in\omega_0\tau} \int dz\, e^{i(K_c-k_z')z}\left(i\frac{K_c}{m}\partial_z+\frac{\nabla^2}{2m}\right)\Phi \label{eq:NLO:amplitude:form}
\end{align}
where we cancelled the opposite phase factors within the temporal integral, applying integration by parts and utilizing the time-periodicity of $\Phi$. Now we proceed perturbatively:
\begin{equation}
    K_c\approx k + \frac{i}{v}\partial_\tau \quad\quad\quad \text{and} \quad\quad\quad K_c-k_z'\approx \frac{q_\perp^2}{2k}.
\end{equation}
With these, (within the temporal integral) 
\begin{align}
    F\approx \frac{-i}{v}\int dz\, e^{iq_\perp^2z/2k}\left(iv\partial_z - \frac{1}{k}\partial_\tau \partial_z+\frac{\nabla^2}{2m}\right)\Phi.
\end{align}
We now treat both $\Phi$ and $e^{iq_\perp^2z/2k}$ perturbatively. In the eikonal approximation, this latter phase factor is set to one. Here, we will retain the linear correction: $e^{iq_\perp^2z/2k}\approx 1+iq_\perp^2z/2k$. Again, we assume this expansion is protected by the potential's envelope at large $z$. The first two terms inside the parentheses, which include all the corrections from the wave function, give
\begin{equation} \label{eq:NLO:z-inelastic}
    iv \left(e^{i\chi}-1\right) + iv\gwf{1}(z\to\infty) - \frac{1}{k}\partial_\tau e^{i\chi}.
\end{equation}

Let's consider their contribution to the NLO observable first, \textit{neglecting} the diffractive correction of the wave function. Subtracting the LO piece, from the form of wave-function corrections (\ref{eq:NLO:gwf}),
\begin{align}
    &{\rm Re}\left(e^{-i\chi}F^{(1)}\right)\supset\frac{-1}{2E}\partial_\tau\chi + \frac{1}{v}{\rm Im}\int_{-\infty}^\infty\Phi^{(0)*}\left[G_{v_n}(z)+G_\phi(z)\right]\ dz \\
    =& \frac{-1}{2E}\left(\partial_\tau\chi-{\rm Im}\int_{-\infty}^\infty\Phi^{(0)*}\partial_\tau\partial_z\gwf{0}(z)\ dz\right) + \frac{1}{v}{\rm Im}\int_{-\infty}^\infty\Phi^{(0)*}D_\phi\gwf{0}(z)\ dz \\
    =&\frac{\varepsilon\qmin}{4v}\int z U\left(\tau+\frac{z}{v}\right)dz = \frac{\varepsilon\qmin}{2v}{\rm Im}\left(e^{i\omega_0\tau}\partial_{\qmin}\zeta\right).
\end{align}
The imaginary part of the integral over $G_{\partial_z^2}$ vanishes. In the second line, the channel velocity correction (\ref{eq:NLO:sources:vn}) cancels with the residual mixed term in the Laplacian. The temporal integral in (\ref{eq:NLO:observable}) is readily evaluated:
\begin{equation} \label{eq:NLO:observable:phi}
    \int_0^{2\pi/\omega_0} \partial_\tau\chi\ {\rm Re}\left(e^{-i\chi}F^{(1)}\right) d\tau \supset \frac{\pi\varepsilon\qmin}{2v^2}\partial_{\qmin}|\zeta|^2.
\end{equation}

We now turn to the remaining diffractive corrections, including the term entering via $\Phi^{(1)}$:
\begin{equation}
    F_{\nabla_\perp^2}^{(1)} = \frac{-i}{2k}\int \left[1-e^{i\chi}\Phi^{(0)*}(z)+z\partial_z\right]\nabla_\perp^2\gwf{0}(z)\ dz.
\end{equation}
Here, we've implicitly applied Green's identity to take $q_\perp^2\to-\nabla_\perp^2$, for the boundary term at $b\to\infty$ vanishes as the potential decays suitably rapidly. If the mediator has a mass, then the asymptotic exponential suppression automatically satisfies this. The unscreened dipole potential also suffices. This simplifies when we compute the observable. Assuming that the potential decays suitably at infinity, or precisely $z\left|\nabla_\perp^2\tilde{\chi}+i\left(\nabla_\perp\tilde{\chi}\right)^2\right|\to 0$ as $z\to-\infty$,
\begin{align}
    {\rm Re}\left(e^{-i\chi}F^{(1)}_{\nabla_\perp^2}\right) &= \frac{1}{2k}\nabla_\perp^2\int z\partial_z\tilde{\chi}\ dz \\
    &= \frac{-2}{4E}\nabla_\perp^2{\rm Im}\left(e^{i\omega_0\tau}\partial_{\qmin}\zeta\right)+\frac{1}{2k}\nabla_\perp^2\int z\partial_z\tilde{\chi}_0\ dz.
\end{align}
The second static term will vanish subsequently, when weighted by $\partial_\tau\chi$ and integrated over time. For the first, within the integral over impact parameters, we may apply Green's identity again
\begin{equation}
    \int d^2\mathbf{b}\ \zeta\partial_{\qmin}\nabla_\perp^2\zeta^* = -\int d^2\mathbf{b}\ \nabla_\perp\zeta\cdot\nabla_\perp\partial_{\qmin}\zeta^*
    \label{eq:byparts}
\end{equation}
where the boundary term vanishes as $\qmin\neq 0$ induces an asymptotic exponential decay of $\zeta$ in $b$. With this,
\begin{equation} \label{eq:NLO:observable:diffrac}
    \int d^2\mathbf{b} \int_0^{2\pi/\omega_0} \partial_\tau\chi\ {\rm Re}\left(e^{-i\chi}F^{(1)}_{\nabla_\perp^2}\right) d\tau = \frac{\pi}{kv^2} \int d^2\mathbf{b}\ \partial_{\qmin}\left|\nabla_\perp\zeta\right|^2.
\end{equation}
Combining these contributions (\ref{eq:NLO:observable:phi}, \ref{eq:NLO:observable:diffrac}) with the kinematic correction in (\ref{eq:NLO:observable}), we obtain at NLO
\begin{align} \label{eq:NLO:observable:position}
    -\sum_{n=-\infty}^\infty n\sigma_n &= \frac{1}{kv^2}\int d^2\mathbf{b}\ \partial_{\qmin} \left(\qmin^2|\zeta|^2 + \left|\nabla_\perp\zeta\right|^2\right)
\end{align}
which graciously agrees with the result arrived at from the quantum description of the target system (\ref{eq:jz}). As may be expected, the time-independent part of the potential $V_0$, which influences elastic scattering, does not enter for this observable. Since $\zeta$ naturally has an asymptotic exponential suppression from $\qmin\neq 0$, we may take $\partial_{\qmin}$ outside the integral. This integral can also be expressed in Fourier space, using
\begin{equation}
    \tilde\zeta(\mathbf{q}_\perp) \equiv  \int d^2\mathbf{b}\, e^{-i\mathbf{q}_\perp\cdot\mathbf{b}}\zeta(\mathbf{b}) = \tilde{V}_-(\mathbf{q}_\perp,q_z=-\qmin) \ ,
\end{equation}
as
\begin{align}
    -\sum_{n=-\infty}^\infty n\sigma_n &=\frac{1}{kv^2}\partial_{\qmin} \int \frac{d^2\mathbf{q}_\perp}{(2\pi)^2} \left(\qmin^2+q_\perp^2\right)\left|\tilde{V}(\mathbf{q}_\perp,-q_0)\right|^2 \\
    &= \frac{1}{kv^2}\partial_{\qmin} \int \left.\frac{d^2\mathbf{q}_\perp}{(2\pi)^2} q^2\left|\tilde{V}_-(-\mathbf{q})\right|^2 \right |_{q_z=\qmin}. \label{eq:NLO:observable:momentum}
\end{align}

\subsubsection{$\mathbf{A}\cdot\hat{\mathbf{p}}$ coupling in momentum-dependent potentials} \label{sec:scattering-formalism:p-dependent potential}

The minimal-coupling prescription for electromagnetic interactions yields a momentum dependent potential term $\propto \{A_i,\hat{p}_i\}$, which conceivably could introduce corrections at NLO. In this section, we show that if the physical potential naturally satisfies $\nabla\cdot\mathbf{A}=0$, as for the dipole potential, or if one first chooses Coulomb gauge, with corresponding modification of the scalar potential, then there is no additional contribution from $\{A_i,p_i\}=2\mathbf{A}\cdot\mathbf{p}$ at NLO to energy transfer, and our result (\ref{eq:NLO:observable:position}) stands.

The effect of the momentum operator in the potential is an additional term in the Schrodinger equation where $\hat{\mathbf{p}}$ acts on the envelope:
\begin{equation*}
    \frac{iq_e}{m} \mathbf{A}\cdot\nabla\varphi 
\end{equation*}
for a charge $q_e$. At LO, this is clearly subdominant to
$$\frac{-q_e}{m} \mathbf{A}\cdot\mathbf{k}\ \varphi$$
which is also contained in the potential. At NLO, this additional term contributes an additional source to the NLO wave function. Without analyzing the wave function into Floquet channels, putting directly $\varphi\approx\varphi^{(0)}+\varphi^{(1)}$, the Schr\"{o}dinger equation becomes, c.f. (\ref{eq:NLO:sources})
\begin{align} \label{eq:NLO:Ap:sources}
    \left[iv\partial_z-V\left(\tau+\frac{z}{v}\right)\right]\varphi^{(1)} = \frac{-1}{2m} \left[\nabla_\perp^2 + \left(\partial_z-\frac{\partial_\tau}{v}\right)^2\right]\varphi^{(0)}+\frac{iq_e}{m}\left[\mathbf{A}_\perp\cdot\nabla_\perp + A_z\left(\partial_z-\frac{\partial_\tau}{v}\right)\right] \varphi^{(0)}.
\end{align}
Here, $V$ encompasses all interactions at LO, including the coupling $\propto \mathbf{A}\cdot\mathbf{k}$. Many of the source terms on the first line of the right coincide with those reported in \ref{app:observable:wave-function}, recalling that $\gwf{0}=\varphi^{(0)}$. The difference is due to the phase transformation connecting $\varphi$ to $\Phi$.

As $V$ is replaced by the full operator $\hat{V}$ in (\ref{eq:NLO:Schro}), which includes $\mathbf{A}\cdot(\mathbf{k}+\hat{\mathbf{p}})$, we see that the amplitude (\ref{eq:NLO:amplitude:form}) remains formally unaffected. Thus, at NLO, we recognize in (\ref{eq:NLO:Ap:sources}) an additional source for the wave-function correction as
\begin{equation}
    G_A\equiv \frac{iq_e}{m}\left[\mathbf{A}_\perp\cdot\nabla_\perp + A_z\left(\partial_z-\frac{\partial_\tau}{v}\right)\right] \gwf{0}.
\end{equation}
Its contribution to the observable is therefore, c.f. (\ref{eq:NLO:observable}, \ref{eq:NLO:z-inelastic}),
\begin{align*}
    {\rm Re}\left(e^{-i\chi}F^{(1)}_A\right) &=  \frac{1}{v}{\rm Im}\int_{-\infty}^\infty\Phi^{(0)*}G_A(z) dz \\
    &= \frac{-q_e}{k} {\rm Im}\int_{-\infty}^\infty \left[\mathbf{A}_\perp\cdot\nabla_\perp + A_z\left(\partial_z-\frac{\partial_\tau}{v}\right)\right]\tilde{\chi}(z) dz \\
    &= 0.
\end{align*}

\section{Spin-dependent scattering}
\label{app:spin_dep}

The spin-spin Hamiltonian is

\begin{equation}
\begin{split}
    H_{\rm int}&=\frac{1}{\Lambda^2}\sum_i
\vec s_P\cdot\vec s_i(t)\,\delta^{(3)}(\mathbf r-\mathbf r_i)\\
&=\frac{n_s(\mathbf{r})}{4\Lambda^2}\pare{s_- e^{i\os t} + s_+ e^{-i\os t}} + \frac{i}{2\Lambda^2}\sum_i\pare{s_- e^{i\os t} - s_+ e^{-i\os t}}s_y^{(i)}\delta^{(3)}(\mathbf{r}-\mathbf{r}_i).
\end{split}
\label{eq:spinspin_internal_sup}
\end{equation}
where $n_s(\mathbf{r})=n_{s,0}\Theta(R-r)$ for a sphere of uniform density $n_s$ and radius $R$.
The semi-classical potential in the first term gives a Schr\"{o}dinger equation for $P$

\begin{align}
    i\frac{\partial\psi}{\partial t}&=-\frac{\nabla^2}{2m}\psi + \frac{n_s(\mathbf{r})}{4\Lambda^2} \pare{\sigma_+e^{-i\omega_0 t}+\sigma_-e^{i\omega_0 t}}\psi,
\end{align}
which can be reduced to a time-independent Schr\"{o}dinger equation by writing

\begin{equation}
    \psi = \frac{1}{\sqrt{V}}\begin{pmatrix}
        e^{-i\os t/2} & 0\\
        0 & e^{i\os t /2}
    \end{pmatrix}\tilde{\psi},
\end{equation}
so that

\begin{equation}
    i\frac{\partial\tilde{\psi}}{\partial t}=-\frac{\nabla^2}{2m}\tilde{\psi} + \parea{\frac{n_s(\mathbf{r})}{4\Lambda^2} \sigma_x-\frac{\os}{2}\sigma_z}\tilde{\psi},
    \label{eq:spin_timeindep}
\end{equation}
where $\sigma_x$ and $\sigma_z$ are the standard Pauli matrices. In this frame $\tilde{\psi}$ has a splitting $\os/2$ everywhere in space, not only within the sphere.

This is a time-independent scattering problem from a spherically symmetric potential and so it can be solved exactly using partial waves. The exact solution is not particularly informative. Instead, because we are interested in particles with $k R\gg 1$ and $V/E_\text{kin}\ll 1$, we can apply an eikonal approximation,

\begin{eqnarray}
    \tilde{\psi}=e^{-i\mathcal{E}t}\begin{pmatrix}
        e^{ik_+z}u_+(\mathbf{b},z)\\
        e^{ik_-z}u_-(\mathbf{b},z)
    \end{pmatrix}=
    e^{-i\mathcal{E}t}\begin{pmatrix}
        e^{ik_+z}&0\\
        0&e^{ik_-z}
    \end{pmatrix}u,
    \label{eq:eik_rotated}
\end{eqnarray}
where $u=(u_+, u_-)^T$ and $k_\pm=\sqrt{2m(\mathcal{E}\pm\omega_0/2)}$. The energy $\mathcal{E}$ is fixed here because Eq.~\eqref{eq:spin_timeindep} is time-independent. In the original frame the two components are split by $\pm\os/2$. Then $u$ satisfies

\begin{equation}
    \frac{\partial u}{\partial z}=-\frac{i n_s(\mathbf{r})}{4\Lambda^2}\begin{pmatrix}
        0 & \frac{1}{v_+}e^{-i\Delta k z}\\
        \frac{1}{v_-}e^{i\Delta k z} & 0
    \end{pmatrix}u,
\end{equation}
where $\Delta k =k_+-k_-\equiv q$. This can be simplified further by setting $u=D(z)w$ such that

\begin{equation}
    u=\begin{pmatrix}
        e^{-iqz/2} & 0\\
        0 & e^{iqz/2}
    \end{pmatrix}w,
\end{equation}
so that

\begin{equation}
    \frac{\partial w}{\partial z} = -i\begin{pmatrix}
        -\frac{q}{2} & \frac{n_s(\mathbf{r})}{4\Lambda^2v_+}\\
        \frac{n_s(\mathbf{r})}{4\Lambda^2v_-} & \frac{q}{2}
    \end{pmatrix}w.
\end{equation}

For $b>R$ the potential is zero and so $u=u_0,$ constant. Similarly, $u=u_0$ for $b<R$ and $z\leq -\ell(b)$, where $\ell(b)=\sqrt{R^2-b^2}$. Inside the sphere $|z|\leq\ell(b)$, $n_s(\mathbf{r})=n_{s,0}$ a constant, and so we have

\begin{equation}
    w_\text{in}(b,z) = 
    \begin{pmatrix}
        \cos\parea{\alpha(z+\ell(b))} + \frac{iq}{2\alpha}\sin\parea{\alpha(z+\ell(b))} & -\frac{ig}{v_+\alpha}\sin\parea{\alpha(z+\ell(b))}\\
        -\frac{ig}{v_-\alpha}\sin\parea{\alpha(z+\ell(b))} & \cos\parea{\alpha(z+\ell(b))} - \frac{iq}{2\alpha}\sin\parea{\alpha(z+\ell(b))}
    \end{pmatrix}
    w_0,\quad |z|\leq\ell(b),
\end{equation}
where $\alpha^2 =\frac{q^2}{4}+\frac{g^2}{v_+v_-}$ and $g=\frac{n_{s,0}}{4\Lambda^2}$ and $w_0 = D(-\ell(b)) u_0$. For $z\geq\ell(b)$, we have $u(b)=D(\ell)w(b,\ell(b))$. 

Taking the incoming $P$ to be asymptotically spin-up, $\psi_\text{in}=e^{-iEt}e^{ikz}(1,0)^T$ at $z\to-\infty$, we have $\mathcal{E}=E-\frac{\os}{2}$ with $E=\frac{k^2}{2m}$ the physical asymptotic incoming energy of the bath. At asymptotic $z\to+\infty$ we then find

\begin{equation}
    \psi_\text{out} = \begin{pmatrix}
        S_{++}(b) e^{ikz}e^{-iEt}\\
        S_{-+}(b)e^{ik_-z}e^{-i(E-\os)t}
    \end{pmatrix},
\end{equation}
with $k_-\approx k-q_0$ and

\begin{align}
    S_{++}(b) &= e^{-iq_0\ell(b)}\parea{\cos\pare{\Omega\frac{\ell(b)}{v}} + \frac{i\os}{\Omega}\sin\pare{\Omega\frac{\ell(b)}{v}}}\\
    S_{+-}(b) &= \frac{-iV_0}{\Omega}\sin\pare{\Omega\frac{\ell(b)}{v}},
\end{align}
where we approximated $v_+\approx v_-$ to leading order and used the notation $\Omega^2=\os^2+V_0^2$ of the main text, for which $g=V_0/2$. By matching to the scattering amplitude at infinity, the cross-section is 

\begin{equation}
    \sigma_\text{flip} = 2\pi\int \di b\, b \abs{S_{+-}(b)}^2,
\end{equation}
which reduces to Eq.~\eqref{eq:sigma_flip}. Taking instead the initial spin to be down, gives the same result, to LO, for a system de-excitation.

The observables can be computed as before. We need the eikonal wavefunctions for fixed asymptotic energy at $-\infty$ in order to expand the many-body operator and compute the response of the spin system to the bath. We find inside the interacting region $b\leq R$ and $|z|\leq \sqrt{R^2-b^2}$ the basis

\begin{align}
    \psi_k^\uparrow &= \frac{e^{i\frac{q}{2}\ell}}{\sqrt{V}}\begin{pmatrix}
        e^{-iEt}e^{ikz}e^{-i\frac{qz}{2}}\pareb{\cos[\alpha(z+\ell(b))] + \frac{iq}{2\alpha}\sin[\alpha(z+\ell(b))]}\\
        \frac{-ig}{v_-\alpha}e^{-i(E-\os)t}e^{ik_- z} e^{i\frac{qz}{2}} \sin[\alpha(z+\ell(b))]
    \end{pmatrix}\\
    \psi_k^\downarrow &= \frac{e^{-i\frac{q}{2}\ell}}{\sqrt{V}}\begin{pmatrix}
        \frac{-ig}{v_-\alpha}e^{-i(E+\os)t}e^{ik_+ z} e^{-i\frac{qz}{2}} \sin[\alpha(z+\ell(b))]\\
        e^{-iEt}e^{ikz}e^{i\frac{qz}{2}}\pareb{\cos[\alpha(z+\ell(b))] + \frac{iq}{2\alpha}\sin[\alpha(z+\ell(b))]}
    \end{pmatrix},
\label{eq:basis_spin}
\end{align}
where the arrows correspond to the initial spinor polarization at $-\infty$ and $E=\frac{k^2}{2m}$ corresponds to its energy.

Energy transfer depends on the expectation value

\begin{equation}
\begin{split}
   I(\mathbf{r})&= -i\avg{\bar{\Psi}(s_-e^{i\os t} - s_+e^{-i\os t})\Psi}_B \\
   &= -i\int\frac{\di^3 k}{(2\pi)^3} \parea{f_\uparrow(\mathbf{k})\pare{\psi_k^\uparrow}^\dagger U(t)\psi_k^\uparrow + f_\downarrow(\mathbf{k})\pare{\psi_k^\downarrow}^\dagger U(t)\psi_k^\downarrow}\\
   &=\frac{g}{\alpha}\int\frac{\di^3 k}{(2\pi)^3} \parea{v_+ f_\uparrow(\mathbf{k}) - v_-f_\downarrow(\mathbf{k})}\sin\parea{2\alpha(z+\ell(b))}
\end{split}
\end{equation}
assuming uncorrelated up/down components, with $U(t)\equiv s_-e^{i\os t} - s_+e^{-i\os t}$, and the distributions $f_{\uparrow(\downarrow)}$ correspond to the phase-space occupation of particles with spin up and down, respectively. Energy transfer is $\frac{\di\avg{J_z}}{\di t}=g\int\di^3 r\, I(\mathbf{r})$, from the master equation, Eq.~\eqref{eq:master}, and 

\begin{equation}
    \int_{-\ell(b)}^{\ell(b)}\di z\, \sin\parea{2\alpha(z+\ell(b))} = \frac{1}{\alpha}\sin^2\parea{2\alpha\ell(b)},
\end{equation}
from which 

\begin{equation}
    \frac{\di\avg{J_z}}{\di t} = \int\frac{\di^3 k}{(2\pi)^3} \parea{v_+f_\uparrow(\mathbf{k})\sigma_\text{flip}^+ - v_-f_\downarrow(\mathbf{k})\sigma_\text{flip}^-},
\end{equation}
with $\sigma_\text{flip}^+\approx\sigma_\text{flip}^-$ to LO. From this it follows immediately that polarized beams retain LO energy transfer, while if the beam is fully unpolarized, then $v_+f_\uparrow(\mathbf{k}) - v_-f_\downarrow(\mathbf{k})\approx2\frac{q_0}{k}f(\mathbf{k})$, that is, suppressed by $q_0/k$ compared to the total rate, as expected. This reproduces the first part of Eq.~\eqref{eq:rates_spin} of the main text.

Decoherence can also be computed directly from the master equation, Eq.~\eqref{eq:master}. Taking $P$ to be a fermion, we have

\begin{equation}
    \frac{1}{4}\sum_{ij}\gamma_y^{ij} = g^2\int\frac{\di^3 k}{(2\pi)^3}\int\frac{\di^3 k'}{(2\pi)^3}\int_R \di^3 r\int_R\di^3 r'\int_0^t\di\tau \sum_{s,s'=\uparrow,\downarrow}f_s(\mathbf{k})\parea{1-f_{s'}(\mathbf{k}')}\pare{\psi_{k,s}^\dagger iU\psi_{k',s'}}_{\mathbf{r},t}\pare{\psi_{k',s'}^\dagger iU\psi_{k,s}}_{\mathbf{r}',t-\tau},
\end{equation}
which should also be averaged against secular terms in $t$. The time integral gives delta functions, which set $k'\sim k$, to leading order in the prefactor for small momentum transfer. The wavefunction products are proportional to $e^{i\mathbf{q}_\perp\cdot(\mathbf{b}-\mathbf{b}')}$, which allows us to use approximation $\di^2\Omega_{k'} \approx k^{-2}\di^2q_\perp$ in the eikonal regime, to apply Parseval's theorem $\int \di^2 q_\perp\int\di^2 b \di^2b' e^{i\mathbf{q}_\perp\cdot(\mathbf{b}-\mathbf{b}')} g(\mathbf{b})g^*(\mathbf{b}') = \int \di b |g(\mathbf{b})|^2$. Using the eikonal basis Eq.~\eqref{eq:basis_spin} to write the wavefunctions, direct algebraic manipulation yields

\begin{equation}
    \frac{\di\avg{J_z^2}}{\di t} = \int\frac{\di^3 k}{(2\pi)^3} \parea{f_\uparrow(\mathbf{k})\sigma_\text{flip}^+ + f_\downarrow(\mathbf{k})\sigma_\text{flip}^-}v,
\end{equation}
which reproduces the second part of Eq.~\eqref{eq:rates_spin} of the main text.

\section{Rate computations}
\label{app:rates}

In this section we compute rates for various applications of our formalism. Section~\ref{app:neutrons} is concerned with thermal neutron beams, Sec.~\ref{app:charged_beam} with finite area beams of charged particles, and Sec.~\ref{app:mcps} with a thermal abundance of millicharge particles.

\subsection{Thermal neutrons}
\label{app:neutrons}
For a unidirectional thermal neutron beam impinging on the target from direction $\hat{\mathbf{n}}$ of temperature $T$ and flux $\Phi$, the phase-space occupation is

\begin{equation}
    f(\mathbf{k}) = \frac{(2\pi)^3\Phi}{2mT^2}e^{-\frac{k^2}{2mT}}\delta^{(2)}(\hat{\mathbf{k}}-\hat{\mathbf{n}}),
\end{equation}
so that $\int\frac{\di^3 k}{(2\pi)^3} f(\mathbf{k}) \frac{k}{m}=\Phi$. The polarized neutron beam rate is then

\begin{equation}
    \Gamma = \frac{\Phi}{2m^2T^2}\int\di k\, k^3 e^{-\frac{k^2}{2mT}}\sigma_\text{flip}(k),
\end{equation}
which we plot in Fig.~\ref{fig:beam_rates} of the main text.

\subsection{Charged monochromatic beams}
\label{app:charged_beam}

We are interested in computing the interaction rate of a monochromatic collimated charged beam with our spin target. Let the beam extent be $B\leq R$, the momentum $k$ and the incoming direction along $z$, where $z$ is the spin-quantization axis here. The potential is

\begin{eqnarray}
    V_+(\mathbf{b},z)=\frac{e}{3}\mathbf{b}\cdot(\mathbf{v}\times\mathbf{M}_+) h(\sqrt{b^2+z^2}),
\end{eqnarray}
where $\mathbf{v}=v\hat{\mathbf{z}}$ and
\begin{eqnarray}
h(r)=
\begin{cases}
1-\pare{\frac{R}{R_c}}^3, & r<R,\\[4pt]
R^3\left(\dfrac1{r^3}-\dfrac1{R_c^3}\right),
& R<r<R_c,\\[7pt]
0,&r>R_c.
\end{cases}
\end{eqnarray}

We can use Parseval's theorem to write the transverse Fourier transform in Eq.~\eqref{eq:jz} as

\begin{equation}
    I = \int\frac{\di^2 q_\perp}{(2\pi)^2}Q^2\abs{\widetilde{V}(\mathbf{Q})}^2 = \int \di^2 b \pare{\abs{\nabla_\perp U_+}^2 + q^2\abs{U_+}^2}
\end{equation}
with 
\begin{eqnarray}
\begin{split}
    U_+(\mathbf{b},q) &= \int\di z\, V_+(\mathbf{b},z) e^{-iqz}\\
    &= \frac{e}{3}\mathbf{b}\cdot(\mathbf{v}\times\mathbf{M}_+) \pareb{2\parea{1-\pare{\frac{R}{R_c}}^3}\frac{\sin q\sqrt{R^2-b^2}}{q} + 2R^3\int_{\sqrt{R^2-b^2}}^{\sqrt{R_c^2-b^2}} \di z\, \cos(qz)\parea{\frac{1}{(b^2+z^2)^{\frac{3}{2}}}-\frac{1}{R_c^3}}},
    \label{eq:uplus}
\end{split}
\end{eqnarray}
for $b\leq R$. A subtlety in deriving Eq.~\eqref{eq:jz} is that boundary terms at infinity were dropped to simplify Laplacian terms, Eq.~\eqref{eq:byparts}. Since we have a finite beam here, these boundary terms induce a $\mathcal{O}(1)$ correction, and so need to be included. The correction to $I$ is 

\begin{equation}
\Delta I = -\oint_{b=B}\di\ell\, \text{Re}\parea{U_-\frac{\partial U_+}{\partial b}}
\end{equation}

Define the term in the curly brackets of Eq.~\eqref{eq:uplus} as $L(b,q)$. Then it can be shown that

\begin{equation}
    I = \frac{2\pi e^2 v^2 |M_+|^2}{9}F_B(q),
\end{equation}
with
\begin{equation}
    F_B(q) \equiv \int b\,\di b\parea{L^2 + b L \frac{\partial L}{\partial b} + \frac{b^2}{2}\pare{\frac{\partial L}{\partial b}}^2+\frac{q^2 b^2}{2}L^2} - \frac{B^2}{2}L\left.\pare{L+B\frac{\partial L}{\partial b}}\right|_{b=B},
\end{equation}
which, again, assumes $B\leq R$. For an incoherent beam the integrand here is proportional to the scattering probability at fixed impact parameter, and so we can directly restrict this integral to $0\leq b\leq B$ to take into account the extend of the beam. Given our assumptions, the rate is

\begin{equation}
    \frac{\di\avg{J_z}}{\di t} = \Phi\frac{2\pi e^2 |M_+|^2}{9 k}\left.\frac{\di F_B(q)}{\di q}\right|_{q=q_0}.
\end{equation}

Similarly for the variance, we can write $
d\Omega_{k'}\simeq\frac{d^2q_\perp}{k^2}$, and use Parseval's theorem again. We have

\begin{equation}
    \frac{\di\text{var}[J_z]}{\di t} = \pi\Phi\int_0^B b\,\di b\, \bar{\chi}^2(b),
\end{equation}
which can also be directly proven by summing $\sum_n n^2\sigma_n^\text{LO}$, and we used the defintion $\bar{\chi}=\frac{2}{v} |U_+|$ of the main text. With the preceding notation, $\bar{\chi}^2=\frac{2e^2|M_+|^2}{9}b^2L^2$, at $q=q_0$. We numerically evaluate the energy transfer and variance rates in Fig.~\ref{fig:beam_rates} of the main text.

\subsection{Thermal millicharged particles}
\label{app:mcps}
In contrast to the discussion above, the accumulated millicharged distribution is isotropic at temperature $T$ with a mass- and charge-dependent density computed in Ref~\cite{Berlin:2025btf}. Using

\begin{equation}
    f(k) = n_\chi\pare{\frac{2\pi}{m_\chi T}}^{3/2}e^{-\frac{k^2}{2mT}},
\end{equation}
we use Eqs.~\eqref{eq:jz} and~\eqref{eq:jz2} to find the thermally averaged rates

\begin{align}
    \frac{\text{d}\left<J_z\right>}{\text{d} t}
    &=
    -
    \frac{2\pi}{27}n_\chi
    \frac{(e q_\chi\gamma n_s)^2 R^5}{m}
    \frac{\os R}{v_T}
    \mathcal{F}\left(\frac{\os R}{v_T}\right),\\
\frac{\di \avg{ J_z^2}}{\di t}
&=
\frac{2\pi}{27}\,
n_\chi
(e q_\chi\gamma n_s)^2
R^6
v_T 
\mathcal F\left(\frac{\os R}{v_T}\right),
\end{align}
where $v_T=\sqrt{2T/m}$, and $\mathcal{F}(\eta)$ is the thermally averaged form factor, 
\begin{equation}
\mathcal F(\eta)
=
\frac{2}{\sqrt{\pi}}
\int_0^\infty\frac{du}{u}
e^{-\eta^2/u^2}
\left[
\frac{3j_1(u)}{u}
-\frac{3j_1(\lambda u)}{\lambda u}
\right]^2
\end{equation}
with $\lambda=R_c/R\gtrsim 1$. For energy transfer, numerically we find the optimal value $\eta_* \mathcal{F}(\eta_*)\approx 0.54$ $\eta_*\approx 0.8$ for $\lambda = 5$. For decoherence, small splittings are favored, $\mathcal F(\eta)\approx \frac{2}{\sqrt{\pi}}\log\lambda$ for $\eta\lambda\ll 1$. We use these formulas in Fig.~\ref{fig:mCP_reach}.

\end{document}